\documentclass[fleqn,usenatbib]{mnras}
\newcommand{\msun}{\ensuremath{M_{\odot}}}
\newcommand{\lsun}{\ensuremath{L_{\odot}}}
\newcommand{\iso}[2]{\ensuremath{^{#1}\rm{#2}}}
\usepackage{newtxtext,newtxmath}

\usepackage[T1]{fontenc}
\usepackage{ae,aecompl}

\usepackage{graphicx}	
\usepackage{amsmath}	
\usepackage{amssymb}	
\usepackage{enumitem}

\title[Ignition of detonation in accreted helium]{Ignition of detonation in accreted helium envelopes
}

\author[Glasner et al.]{
	 S. Ami Glasner,$^{1}$  	E. Livne ,$^{1}$,
	E. Steinberg,$^{1,2}$ A. Yalinewich,$^{1,3}$   James W. Truran,$^{4}$
	\\
$^{1}$Racah Institute of Physics, The Hebrew University, Jerusalem 91904,  Israel\\
$^{2}$ Columbia Astrophysics Laboratory, Columbia University, 550 West 120th Street, New York, NY 10027, USA \\
$^{3}$ Canadian Institute of Theoretical astrophysics, 60 St. George St., Toronto. ON M5S 3H8, Canada\\
$^{4}$ Department of Astronomy and Astrophysics, University of
Chicago, 5640 South Ellis Avenue, Chicago, IL 60637
}

\date{Accepted 2018 February 6. Received 2018 January 26; in original form 2017 November 29} 
\pubyear{2015}

\begin{document}
\label{firstpage}
\pagerange{\pageref{firstpage}--\pageref{lastpage}}
\maketitle

\begin{abstract}

Sub-Chandrasekhar CO white dwarfs accreting helium have been considered as candidates for Type Ia supernova(SNIa) progenitors since the early 1980s (helium shell mass $> 0.1 \msun $). These models, once detonated did not fit the observed spectra and light curve of typical SNIa observations. New theoretical work examined detonations on much less massive ($< 0.05 \msun $) envelopes.  They find stable detonations that lead to light curves, spectra and abundances that compare relatively well with the observational data. The exact mechanism leading to the ignition of helium detonation is a key issue, since it is a mandatory first step for the whole scenario. As the flow of the accreted envelope is unstable to convection long before any hydrodynamic phenomena develops, a multidimensional approach is needed in order to study the ignition process. The complex convective reactive flow is challenging to any hydrodynamical solver. 
To the best of our knowledge, all previous 2D studies ignited the detonation artificially. We present here, for the first time, fully consistent results from two hydrodynamical 2D solvers that adopt two independent accurate schemes. For both solvers an effort was made to overcome the problematics raised by the finite resolution and numerical diffusion by the advective terms. Our best models lead to the ignition of a detonation in a convective cell.  
Our results are robust and the agreement between the two different numerical approaches is very good.   
 
\end{abstract}




\section{Introduction}
   
 For helium accretion rates around  $ ~10^{-8} $ \msun/yr on top of a massive CO white dwarf (WD), a thick helium layer would accumulate and ignite ($> 0.1$ \msun) \citep{Taam80, Nomoto80, Nomoto82, Woosley86}. The possible outcomes are as follows:        

- Helium runaway ignited as a single detonation or a subsonic burning flame, producing a faint SN and leaving behind an intact WD. 
  
    - The ignited helium detonation leads to a secondary explosion of the CO core as well as disruption of the whole star \citep{Taam80, Nomoto80, Nomoto82, Woosley86,Livne90, LivGla90, LivGla91, WoosleyWeaver94}.

 For this scenario there are two possible channels:

   a) direct ignition of an inward detonation at the interface between the detonated helium and the CO core, and
   
    b) converging compression wave that steepens as it converges to the inner parts of the WD core and develops to a detonation.    

  Theoretical spectra and light curves are not consistent with those of normal Type Ia supernovae \citep{HoeflichKhokhlov96, Nugent97, Garc99} due to over production of iron group elements. New observations and theoretical work examined cases for which a detonation is assumed to be ignited on much less massive ($< 0.1$\msun) envelopes.  They find out stable detonations that lead to light curves, spectra and abundances that compare relatively well with observational data \citep{Bildsten07, Shen10, Kromer10, WoosleyKasen11, Poznanski10, Kasliwal10, Perets11, Waldman11, Drout13, Insera14}. With a lower helium shell mass, there is no longer a significant over-production of Fe-peak elements at high velocities, which brings model spectra into better agreement with SN Ia observations.   
  
  The exact mechanism leading to the ignition of helium detonation is a key issue since it is a mandatory first step for the whole scenario.  Evolutionary one-dimensional (1D) models show that the accreted helium envelope is unstable to convection on time-scales of days, prior to the runaway.   Examination of the relevant time-scales shows that at that stage the time-scale hierarchy is    $\tau_{hyd} < \tau_{conv} < \tau_{burn} $ [where $\tau_{hyd}$ is the dynamical time-scale (pressure scaleheight/speed of sound),  $\tau_{conv}$  is the convective turnover time (derived from the 1D mixing length theory, \cite{spi63}) and $\tau_{burn}$ is the time-scale for energy release by helium burning (the time in which isobaric burning increases the temperature up to $1.0\times10^{9}K$)]. As long as this is the order of time-scales, the structure of the burning envelope is almost spherical symmetric. The energy release by the burning of helium increases the temperature and as long as the matter stays partially degenerate the changes in the pressure profile are negligible. Since the temperature continues to rise there comes a stage (temperature $~2.5\times10^{8}K$) where the increased burning rate shortens dramatically the burning time, $\tau_{burn}$,   and the time-scale hierarchy becomes $\tau_{hyd} < \tau_{burn} < \tau_{conv} $. From there on one can expect that local temperature fluctuations can lead to further decrease in the burning time so that in a local region that is big enough the time-scales are $\tau_{burn} < \tau_{hyd} $ and a detonation can be ignited locally. In order to study such fluctuations and their ability to ignite a detonation a multidimensional approach is needed.   Most of the multidimensional numerical models published to date ignited the detonation artificially and ignored the convective flow (\cite{Fink10}; \cite{Townsley12}; \cite{MollWoosley13}; \cite{Sim12}). 
  
  The ignition of the helium detonation was studied mainly by 1D models.  Early work by \cite{Nomoto82} and \cite{WoosleyKasen11}  analysed the problem with crude resolution.   Lately \cite{Holcomb13}  and \cite{ShenMoore14} made detailed 1D parametric studies in order to find the conditions (critical size, density and temperature profile) for the ignition of a helium detonation by the Zeldovich mechanism  (\cite{Zel70}; \cite{BlinnikovKhokhlov87}). Shallow temperature gradients are necessary to initiate a   detonation by the Zeldovich mechanism; steep ones inhibit it. Convection, so long as it operates efficiently, keeps the temperature gradient adiabatic. Defining the burning phase velocity to be the ratio between the distance of two points and the difference in their burning time, $\tau_{burn}$, a detonation can occur if a big enough region exists that the burning phase velocity is supersonic. On the other hand, if the burning phase velocity is subsonic, a subsonic flame results.   In a series of papers \cite{Zingale13} and \cite{Jacobs16} made a comprehensive study in order to characterize the early stages of the reactive convective flow of helium using a multidimensional implicit low mach number solver. In the low mach number solver approximation sound waves are filtered out of the system, but compressibility effects
due to stratification and local heat release are retained. Therefore, such schemes can follow the evolution of the convective flow as long as the dynamic time-scale is the slowest scale. 

    In our study we solve the full 2D hydrodynamic equations in order to resolve the dynamic stages of the reactive convective flow once the burning time-scale becomes the shortest scale. Due to the extreme temperature sensitivity of the burning rates, the spatial scales involved in the ignition process range from a microscopic scale of a few centimetres to the macroscopic scales of the scaleheight in the star. With the existing hydrodynamical solvers it is impossible to resolve the entire range of scales. The challenge of any numerical research under those constraints is to find significant physical results with the best achievable spatial resolution. The main challenge current numerical schemes have is to minimize artificial enhancement due to the finite resolution and due to numerical diffusion.
    In Eulerian simulations the combination of incomplete resolution and advective terms can lead to numerical heating by hot ashes from adjacent burning regions, heating that can artificially enhance the local burning and develop into a `numeric' detonation. 
     
      We present here, for the first time, fully consistent results from two hydrodynamical 2D solvers that adopt two independent accurate schemes, \textsc{vulcan} \cite{Livne93} and \textsc{rich} \citep{RICH15}. For both solvers, an effort was made to overcome the problematics raised by the finite resolution and numerical diffusion by the advective terms. Our best models lead to the ignition of a detonation within the inflow of a convective cell. Although our models are only in 2D, we expect that more realistic  models in 3D will have  higher temperature fluctuations with higher burning rates. Therefore, 3D models will not alter our major findings concerning the ignition of a  detonation.
Our results are robust and the agreement between the two different numerical approaches is very good.
      
      This paper is organized as follows. In Section 2, we describe the two solvers that we use to simulate the convective reactive flow.
The initial set-up is presented in Section 3. In Section 4, we describe the main results of our simulations, and in Section 5, we give a brief discussion and summary.
 
 \section{Transition to Detonation in Numerical Simulations}
 
 As stated above, numerical simulations of reactive flow may easily give wrong results when the burning rates are high. Especially, artificial transition to detonation appears in both Lagrangian and Eulerian simulations. In 1D Lagrangian simulations the problem occurs when the local burning characteristic time $t_{b}=E_{th}/\dot{E}$ is shorter than the zone's dynamical time $t_{s}=\delta{x}/c$, where $E_{th}$ is the internal energy per gram, $\dot{E}$ is the rate of energy released by burning and $c$ is the local speed of sound.
 In typical astrophysical problems, like burning in WDs, the burning rates become so fast that any reasonable spatial resolution cannot satisfy the condition $t_{b}>t_{s}$. In those cases burning rates rise exponentially above some 'critical' temperature, and then numerical instability may lead to spurious detonation. A simple burning limiter can be used to control the burning rates without destroying the dynamical evolution. It is enough to limit the burning rate by $\dot{E}_{lim}=min(\dot{E},f E_{th}/t_{s})$, where $f$ is a numerical factor less than unity. In  \cite{Kush13} it is shown, by examples,
 that the limiter becomes less and less important with increasing resolution, and that the numerical result, concerning the ignition of a detonation, converges rather fast to the correct Lagrangian result. We adopt this limiter in all of our simulations.
 
 The problem is much more severe in multidimensional simulations that, as stated above, must be performed using Eulerian schemes. In that case, advective mixing of fuel and ashes introduces another source for artificial enhancement of burning. For conditions typical to WDs, even a tiny amount of mixing ashes into fresh fuel can excite fast burning if the fuel temperature is close to the critical temperature. So far, there is no robust and accurate method that solves this problem. We have explored this 'mixing problem' and demonstrated artificial transition to detonation in our helium burning problem. As ad hoc remedies we used two independent strategies:
 first, to separate fuel and ashes using `multi-phase' technique in \textsc{v2d} (vulcan-2d; \cite{Livne93} as described in the following subsection. Secondly, to use the `Nearly Lagrangian' solver \textsc{rich} \cite{RICH15}, which minimizes the advection terms compared to \textsc{V2D}. This second technique is described in another subsection.

\begin{figure}

	\includegraphics[width=\columnwidth]{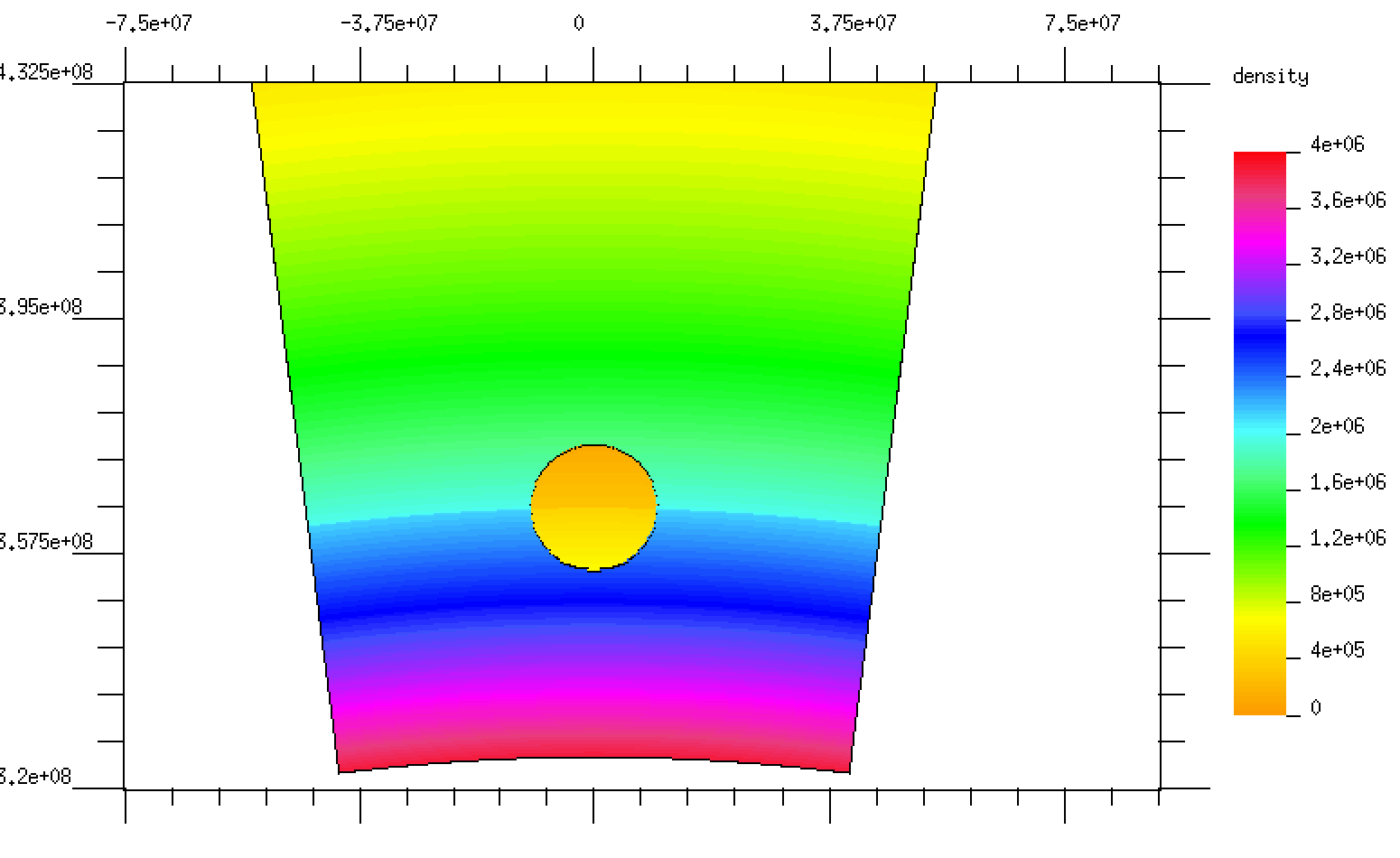}
	\caption{  Density colour map of the in initial  state for the floating torus.}
	\label{fig:bulb_0}
\end{figure}

\begin{figure}
	
	\includegraphics[width=\columnwidth]{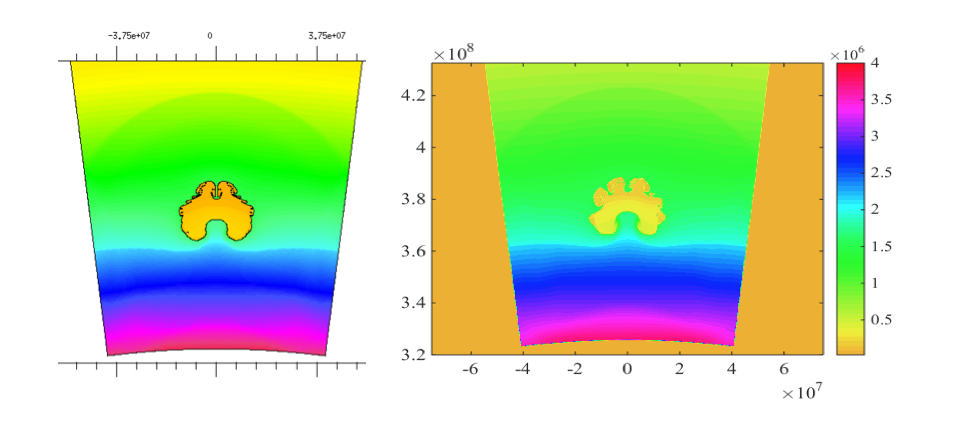}
	\caption{  Density colour map of the floating torus at time 0.2 sec (left-hand panel: \textsc{vulcan}; right-hand panel: \textsc{rich}.}
	\label{fig:bulb_02}
\end{figure}

\begin{figure}
	
	\includegraphics[width=\columnwidth]{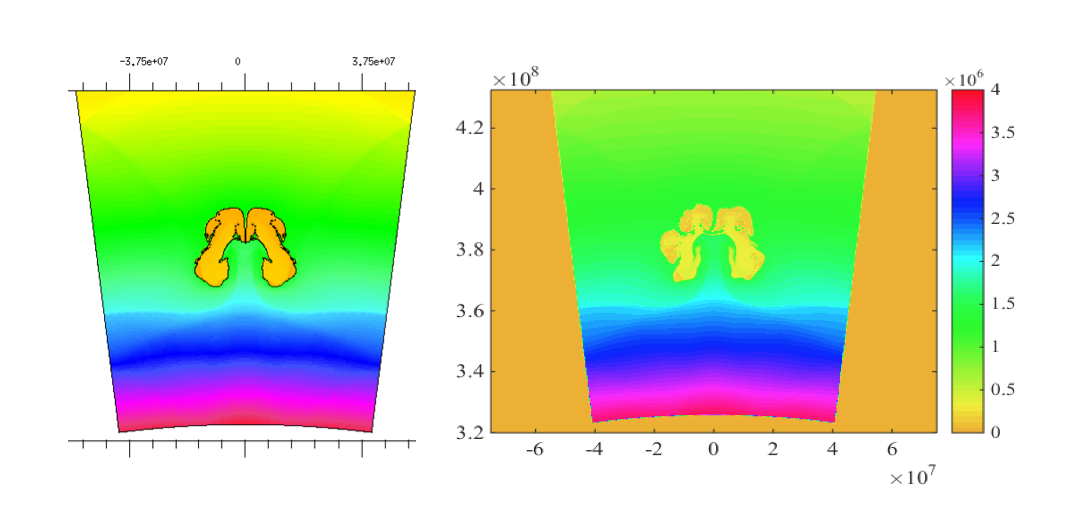}
	\caption{  Density colour map of the floating torus at time 0.3 sec (left-hand panel: \textsc{vulcan}; right-hand panel \textsc{rich}.}
	\label{fig:bulb_03}
\end{figure}

 We have tested the two solvers on non-reactive problems where a buoyant torus with a radius of 100 km and temperature of $~2.0 \times 10^{9} \ K$, initially at hydrostatic equilibrium, floats upwards (Fig.~\ref{fig:bulb_0}). For the floating test, we found good agreement between the two solvers. In Figs \ref{fig:bulb_02} and \ref{fig:bulb_03}, we present the floating torus at times 0.2 and 0.3 s (left-hand panel: \textsc{vulcan}; right-hand panel \textsc{rich}).
 For the helium burning problem, discussed in this study, we find global agreement between the two solvers, whereas there are larger differences in the small-scale features. 
 
 In particular, although the two solvers successfully suppress numerical transition to detonation in very different ways, the final conclusions are the same.
 
 Finally, some caution must be taken due to the issue of possible turbulence. We recognize that our simulations cannot follow fully developed turbulence since
 the spatial resolution is too coarse and also the missing third dimension limits the correct cascading of turbulent scales. It is well known that fully developed
 turbulence may enhance burning and as a result may trigger detonations. However, since we reach detonations in our 2D simulations, the presence of possible turbulence cannot alter our final conclusions.

 \subsection {The \textsc{vulcan} methodology}
 \label{sec:vul}
 
The code \textsc{v2d} (Vulcan-2D; \cite{Livne93}) consists of an ALE method with a second-order advection scheme. In the problems discussed here, the grid moves rather slowly
and the simulations are nearly Eulerian. For astrophysical combustion problems,
a `multi-phase' method has been added, which allows to handle separately fuel and ashes as different substances. The advection
algorithm can use either a `slopes method' (a la Van-Leer scheme) or VOF (Volume of Fluid) method. VOF is actually a family of schemes in which sharp
boundaries between phases are constructed for advecting phases with minimal numerical diffusion \citep{Hirt81,Ubbnik99,Gopala08}.
In \textsc{v2d}, we implement the basic idea during the advection step,
by constructing , in each mixed cell, a local planar interface using the known partial volume fractions. For two phases, the partial volumes
and the orientation of the surface provide  enough information for interface reconstruction. The orientation, or a vector normal to the interface,
is calculated from the phases gradients, based on cell centres values. In cases of
more than two phases, the algorithm is much more complicated, not uniquely defined and less accurate. Once we have the local linear interface, 
we use it to cut the boundaries of the cell and to compute exact phase fluxes between cells. The VOF method is rather accurate as long as the boundaries
between the phases are smooth. When small-scale fragmentation develops it is harder to define accurate normals and as a result the method becomes less reliable.

The VOF method is more appropriate to the issue of this work and was implemented in all our helium burning simulations. 
The phases are defined according to a `critical temperature' which is in our case $T_{c}=3 \cdot 10^8$K. 
Namely, any mass element with $T \le T_{c}$ is defined as phase no. 1, and hotter gas is defined as phase no.2. As long as the phases can be separated, there is
only slight artificial enhancement of burning due to advective mixing. In principle, we would like to create more phases by defining more temperature bins, but when
the number of phases exceeds two, the VOF algorithm becomes much more complicated and less accurate.

 \subsection {The \textsc{rich} methodology}
   Voronoi based moving mesh codes such as \textsc{rich} \citep{RICH15}
    and \textsc{arepo} \citep{arepo},
 which are semi-Lagrangian, greatly reduce mixing by advection terms when solving the Euler equations. 
   In these schemes, cells are constructed via a Voronoi diagram from a set of mesh generating points. A velocity is then assigned to each mesh generating point, which then translates to a velocity of the interfaces between cells. The Riemann problem is then solved in this moving reference frame and the rest follows as in any other finite volume scheme. We refer the reader to \cite{arepo} for more details.
   In some cases, it is enough to advect less than a percent of the cell's mass in order to give rise to spurious detonations, which is less than that is typically advected in \textsc{rich}. For example, advecting $1\%$ (mass fraction) from a cell with $T=2\cdot10^9$K to a cell with $T=2.5\cdot10^8$K (assuming a density $\rho=2\cdot10^6\;\textrm{g/cm}^{3}$ for the cold cell and that the cells are in pressure equilibrium) increases the temperature of the cold cell to $3.56\cdot10^8$K and raises the burning rate by a factor of 66.

   We utilize a two-pronged approach to prevent non-physical detonations. 
   The first modification is a new prescription on how to move the Voronoi mesh generating points that minimizes advection.
   The second change prevents artificial detonations by eliminating the advection term in the fluxes.
 \subsubsection{Improved Mesh Steering}
 \label{sec:point_motion}
 Typically, the velocity that is assigned to a Voronoi mesh generating point is the fluid's velocity \citep{arepo}. For smooth flows, this velocity greatly reduces the advected flux since the interfaces between cells move with a velocity that is close to the velocity of the contact discontinuity. However, when the flow is non-smooth, the advection term is not minimized. 
 We propose a new method for calculating the velocity of the mesh generating point that is based on the actual solution of the Riemann problem. 
 
 We start by assigning each mesh generating point a velocity that is equal to the fluid's velocity. For each interface, labeled $i$, we solve its associated Riemann problem and extract the velocity of the contact discontinuity, $\vec{w}_i$, and the interface's length/area, $L_i$ (2D/3D).
 Our goal is to minimize the differences between all $\vec{w}_i$ and the velocities of the interfaces. Since the velocity of the interface is determined by the velocities of its two neighbouring mesh generating points, we employ an iterative scheme, using typically 5-10 iterations. Each iteration, $n$, consists of updating the current mesh generating velocity of a given point $\vec{v}^n$ in the following fashion
\begin{equation} 
 \vec{v}^{n+1}=\vec{v}^n+\alpha\frac{\sum_i \hat{i}L_i(\vec{w}_i-\vec{v}^n)\cdot\hat{i}}{\sum_i L_i} 
\label{eq:move}  
 \end{equation}

 where $\hat{i}$ is a unit vector normal to the interface (pointing outwards), the summation is done over all of the cell's neighbours and $\alpha$ is typically chosen to be 0.25.
 If there are large density ratios between neighbouring cells, one can optionally add a weight to the right-hand term in equation \ref{eq:move} that is the ratio of the densities.
 \subsubsection{Lagrangian Voronoi Hydrodynamics}
Although minimizing the mass flux between cells helps prevent most spurious detonations, it does not prevent all of them.
 One approximation that can be done is to solve the Riemann problem on an interface as if it was moving with the velocity of the contact discontinuity, $\vec{w}_i$ [this resembles the MFM method in GIZMO \citep{gizmo}, where no mass is advected between cells by assuming the interface moves with the contact discontinuity]. This ensures that there is no mass flux, but gives an error in the hydrodynamical solution (although still a conserving scheme). For smooth flows, the difference between the interface velocity, $\vec{W}_i$, and $\vec{w}_i$, scales inversely with the resolution and as a consequence so does the relative error in the flux.
 
 An even better approximation can be obtained if we deviate from the conserving nature of finite volume schemes.
 The error that arises from the difference between velocity of the interface and $\vec{w}_i$ can be treated not as an error in the flux calculation, but rather as an error in the volume of the cell. The error in the cell's volume is given by 
 \begin{equation}
 dV_i=L_i(\vec{w}_i-\vec{W}_i) \Delta t\hat{i}
 \end{equation}
 where $\Delta t$ is the timestep.
 Since it is impossible to correct the cell's volume while ensuring a proper Voronoi tessellation, we modify the cell's extensive variables, $\mathbf{U}$, instead.
 After $\mathbf{U}$ is updated with the hydrodynamical fluxes, we modify it to be
 \begin{equation}
 \tilde{\mathbf{U}}=\mathbf{U}\frac{V}{V+\sum_i dV_i}
 \label{eq:uu}
 \end{equation}
 where $V$ is the cell's volume at the end of the time-step. This modification ensures that the cell has the correct primitive variables at the expense of non-conservation of the conserved quantities.
 For smooth flows, this leads to a non-conservation that scales inversely with the resolution squared.
 
 \subsubsection{Implementation for this work}
 In light of the discussion above, we implement the following scheme for solving the Euler equations.
 The velocity of the mesh generating points is set by the scheme described in Section \ref{sec:point_motion}.
 
 We divide the cells into two populations, a cold population with temperature less than $3.0\cdot 10^8 \ K $ and a hot population of cells that are hotter. 
 This division arises from equating the time it takes nuclear burning to double a cells temperature to the turn over time of the large eddies that are produced. We also tried setting a temperature of $2.5\cdot10^8 \ K$ for the division between cold and hot cells and got qualitatively similar results.
 
 For interfaces between two cold cells or two hot cells, we calculate the flux with no additional corrections.
 If the interface is between a hot cell and a cold cell, we apply the non-conserving modification. This ensures that the cold material does not ignite due to mixing with hot material.
 
 This leads to a non-conservation of the cold material that is small. During the entire run of the simulation (including the violent burning phase), the error in the mass conservation is less than $3\%$. 
 Since typically the velocity of the interface is slower than the contact discontinuity velocity, the non-conservation leads to a decrease in the mass of the hot gas while increasing the mass of the cold gas. This makes the transition  to detonation even more robust.   

In order to check the affect of the non conservation, we run an additional simulation with the same initial conditions but with a modified hydro scheme.
We follow the same procedure as described above in equation \ref{eq:uu}, but if the fractional difference in volume, $dV_i$ is
smaller than $3\cdot10^{-4}$, we set $dV_i$ to be zero, since the primitive variables hardly change from the correction that arises from this interface. 
Additionally, we cap the maximum fractional value of $dV_i$ to be 0.01 to prevent rare, large, non conservations. 
With those changes, the non conservation in the mass is reduced
to be less than a percent, whereas the overall dynamical evolution and transition from deflagration to detonation are qualitatively the same.

\section{The initial model methodology}

\subsection{The 1D solver}
  The accretion process and the evolution approaching the runaway are studied with a 1D Lagrangian implicit code that integrates the equations of momentum and energy
  conservation assuming spherical symmetry. The energy transfer  includes
  a radiative term and a convective term. The radiative component is
  a diffusive flux. The radiative diffusion coefficient is determined
  according to  the Iglesias \& Rogers opacity tables \citep{IGR96} for
  temperatures above one million Kelvin and according to the Alexander fit
  for lower temperatures \citep{Alexander94}. Electron conductive opacities are computed according to the Itoh fit \citep{Itoh83}. The convective energy flux is computed
  according to the mixing length theory \citep{spi63}. Within a convective
  zone the matter is mixed using a diffusion coefficient that takes into
  account both the local convective velocity and the mixing length
  (which is taken to be two pressure scaleheights):
\begin{equation}                                                                                                                                                    
  D_{c} = v_{c}l/3                                                                                                                                      
\end{equation} 
  
  where $D_{c}$ is the diffusion coefficient,
  $v_{c}$ is the 1D convective velocity, derived from the mixing
  length theory, and $l$ is the effective mixing length.
  
  In the present survey, we ignore element settlement by diffusion and
  undershoot mixing. The equation of state employed to derive the energy
  and pressure for given temperature density and chemical abundances is
  an ideal gas for the ions and involves a table fit for the numerical
  integration of the Fermi integrals for the electrons (and positrons).
  The rate of nuclear energy production and the abundance changes are
  calculated with a 13 nuclei alpha network. The elements in the net are:
  
  \iso{4}He, \iso{12}C, \iso{16}O,\iso{20}Ne, \iso{24}Mg,\iso{28}Si, \iso{32}S,\iso{36}Ar, \iso{40}Ca,\iso{44}Ti, \iso{48}Cr, \iso{52}Fe, and \iso{56}Ni.
  
\begin{figure}
 	\includegraphics[width=\columnwidth]{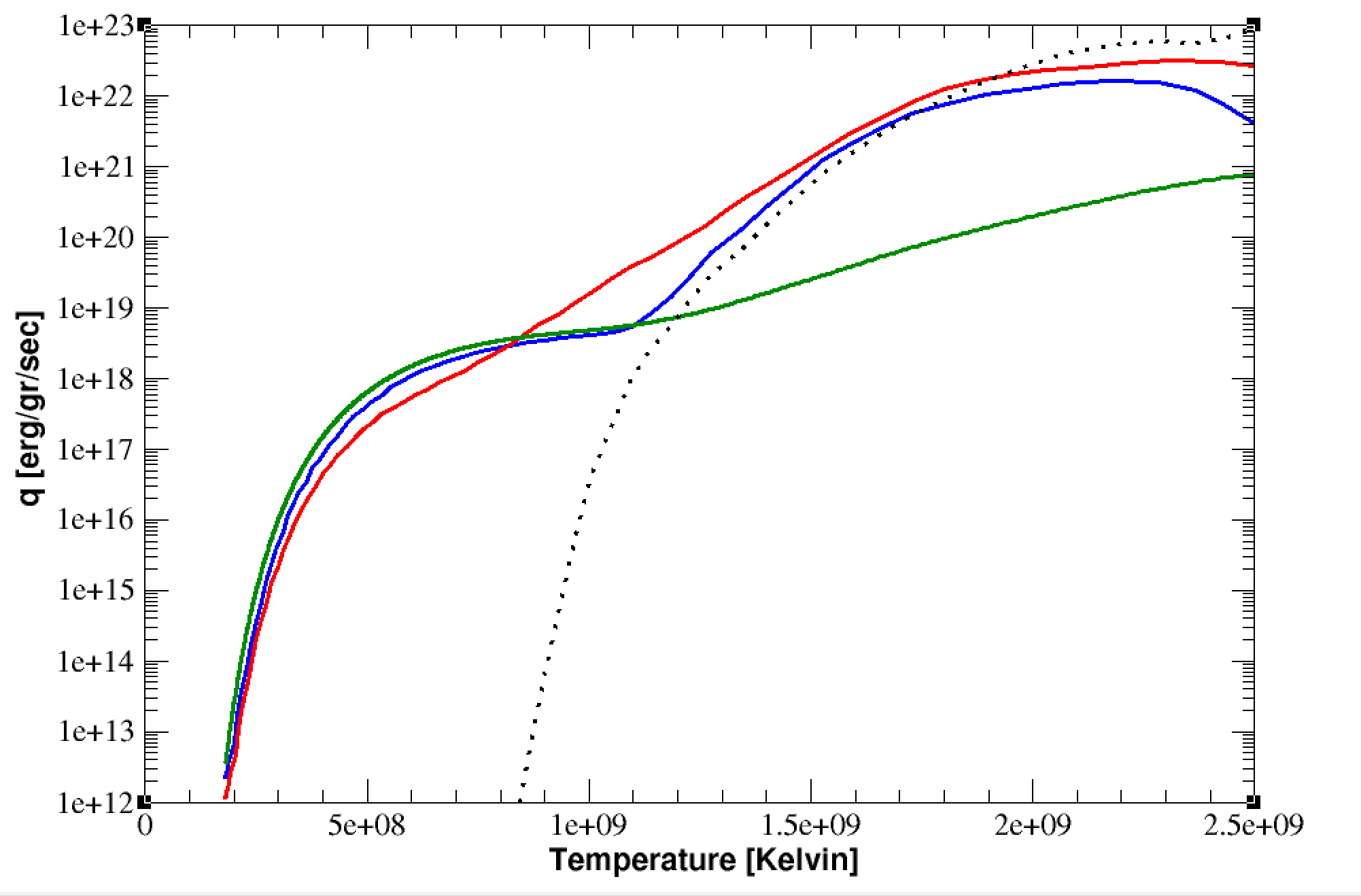}
 	\caption{  Helium burning the energy production rates. Green - 13 elements alpha net, blue - full net 216 elements (both for pure helium), red - full net 216 elements (mixture of  80 \% helium 10 \% carbon and 10 \% oxygen). The dotted line shows the proton mass fraction multiplied by $10^{26}$}
 	\label{fig:helium_rates}
 \end{figure}

The reaction sequence
\iso{12}C (p, $\gamma$ ) \iso{13}N ( $\alpha$ ,p) \iso{16}O 
plays a critical role in accelerating the
burning during the runaway \citep{ShenBild09, WoosleyKasen11}. Yet, as can be seen in Fig.~\ref{fig:helium_rates}, the enhancement is only for temperatures above $~1.0 \times 10^{9} \ K$. Enhancement at that temperature can effect the burning at late stages, as we will discuss later but it does not effect the burning time, $\tau_{burn}$, at the initial low temperature $~2.5 \times 10^{8} \ K$.
We compared full 1D models with the 13 elements alpha net and models with a full net of 216 elements that includes all the relevant isotopes from protons to \iso{70}Zn.  For both cases we indeed got the same results concerning temperature profiles and burning times up to the stage where the maximal temperature was $~1.0 \times 10^{9} \ K$. 

  Accreted material is added each time-step to the outermost zone as
  dictated by the accretion rate and the size of the time step. The
  outermost zone in the numerical grid  is divided into two grid zones
  whenever the mass of the zone is greater than a specified value
  (taken to be twice the specified mass resolution in the envelope).
  The mass is added with the instantaneous thermodynamic properties of
  the outermost grid zone.
  
  \subsection{The details of the initial models}
  The initial model consists of a 1D, spherical symmetric, $1.0 \msun$ CO WD in hydrostatic and thermal equilibrium, cooled to the stage at which the luminosity of the WD is about $ 0.1 \lsun $ (Central temperature of $6 \times 10^{7} \ K$). Using the accretion module of the 1D hydro-evolution solver, helium is accreted on to the surface of the CO core continuously.
  As helium is accumulated, the pressure and the temperature at the base of
  the envelope increase. Once the temperature at the bottom reaches $1 \times 10^{8} \, K$, the heat produced by nuclear burning cannot be removed any more by thermal conductivity and the envelope becomes unstable to convection.
  Since the mass is still increasing, the temperature continues to grow and nuclear reaction rates are enhanced.
   In order to check the sensitivity of our results to the mass of the accreted envelope we calculated two models with different accretion rates.
   
   For all the computed models (Table \ref{tab:models}), a WD of $1.0 \msun$ accretes helium. The type A models are those that accrete $0.1 \msun$ of helium at the rate of $0.86 \times 10^{-8} \msun$  $~ yr^{-1}$ up to the runaway. The type B models accrete $0.05 \msun$ of helium at the rate of $2.00 \times 10^{-8} \msun$  $~ yr^{-1}$ up to the runaway.
   
   \begin{table}
   	\centering
   	\caption{Details of the computed models.}
   	\label{tab:models}
   	\begin{tabular}{lcccccr} 
   		\hline
   		Model & M-CO  & M-Helium &   Angle  & Resolution [km] & Comment &\\
   		& \msun &  \msun   &(radians) & (rad. X trans.) &         &\\
   		\hline
   		A  & 0.186 & 0.10 & 0.08 & 1.6X2.4  & \textsc{vulcan} \\
   		A  & 0.186 & 0.10 & 0.08 & 1.5      & \textsc{rich} \\
   		BW & 0.170 & 0.05 & 0.20 & 1.6X2.4  & \textsc{vulcan} \\
   		B  & 0.086 & 0.05 & 0.08 & 1.5      & \textsc{rich}   \\
   		AW & 0.186 & 0.10 & 0.20 & 1.6X2.4  & \textsc{vulcan} \\
   		AD & 0.178 & 0.10 & 0.12 & 0.95X1.3 & \textsc{vulcan} \\
   		\hline
   	\end{tabular}
   \end{table}

\begin{figure}

  	\includegraphics[width=\columnwidth]{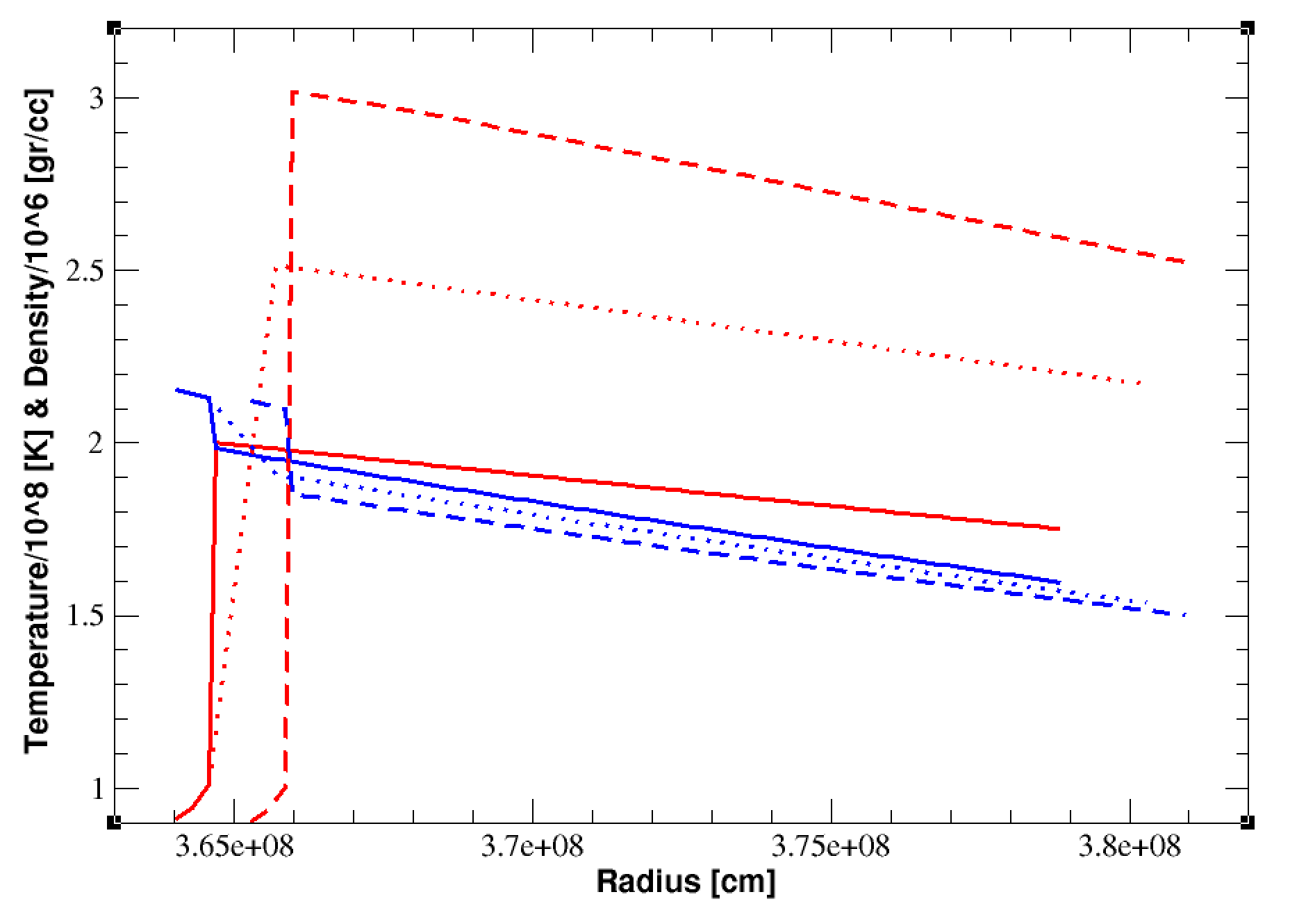}
  	\caption{Temperature profiles (red in units of $1 \times 10^{8} \ K$ ) and density profiles (blue in units of $1 \times 10^{6} \ gr/cc $)  in the helium envelope. For: $Temp_{base}=2 \times 10^{8} \ K$ full line , $Temp_{base}=2.5 \times 10^{8} \ K$ dots, and $Temp_{base}=3 \times 10^{8} \ K$ broken line}
  	\label{fig:tmp_rho}
  \end{figure}

\begin{figure}
 	
 	\includegraphics[width=\columnwidth]{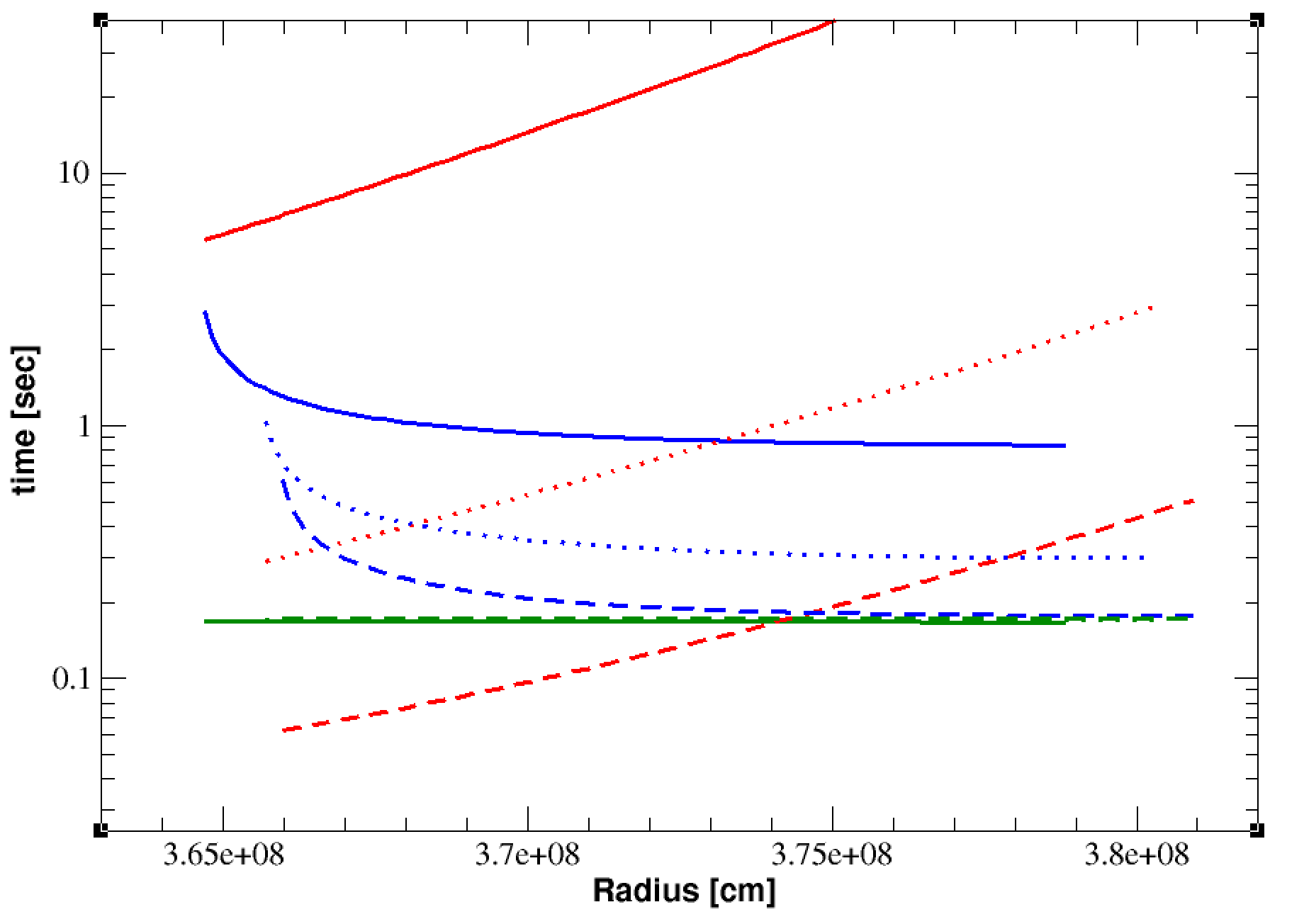}
 	\caption{  The time-scales $\tau_{burn}$ (red) , $\tau_{conv}$ (blue),  $\tau_{hyd}$ (green). $Temp_{base}=2 \times 10^{8} \ K$ full line , $Temp_{base}=2.5 \times 10^{8} \ K$ dots, and $Temp_{base}=3 \times 10^{8} \ K$ broken line}
 	\label{fig:tau}
 \end{figure}

   For the evolutionary stages shown in Fig.~\ref{fig:tmp_rho},  we present in Fig.~\ref{fig:tau} the competing time-scales for models of type A: $\tau_{burn}$, $\tau_{conv}$, and $\tau_{hyd}$. As long as the temperature at the base of the envelope ($Temp_{base}$) is lower than $~2.5 \times 10^{8} \ K$ (Fig.~\ref{fig:tau}), the time-scale hierarchy is $\tau_{hyd} < \tau_{conv} < \tau_{burn}$, and the 1D approach is physically justified. Once the burning time is shorter than the convective turnover time local structure in each convective cell cannot be ignored and the 1D approach is not justified any more \cite{ WoosleyKasen11}.
  
  In the 1D model, it takes 20s for the temperature to rise from  $~2.0 \times 10^{8} \ K$ to $~3.0 \times 10^{8} \ K$, whereas the convective turnover time at these stages is decreasing starting from 2s. Therefore, in order to enable the 2D models to develop a stable convective flow prior to the stage when runaway conditions are met, we map the 1D model into 2D when the base temperature is   $~2.0 \times 10^{8} \ K$.

  The later stages of the reactive flow are studied by the 2D solvers. Both 2D hydrodynamic solvers preserve hydrostatic equilibrium, accurate to better than
  one part in ten thousand. In the next section we present the 2D reactive flow. Although both models ignite the helium and share many qualitative features the ignition itself is robust for the models with accreted helium mass of $0.10 \msun$ (models of type A in Table~\ref{tab:models}) and marginal for the models with accreted helium mass of $0.05 \msun$ (models of type B in Table~\ref{tab:models}). 
  
\section{The 2D reactive flow}

\subsection{The evolution to ignition of models of type A}

\begin{figure}
	
	\includegraphics[width=\columnwidth]{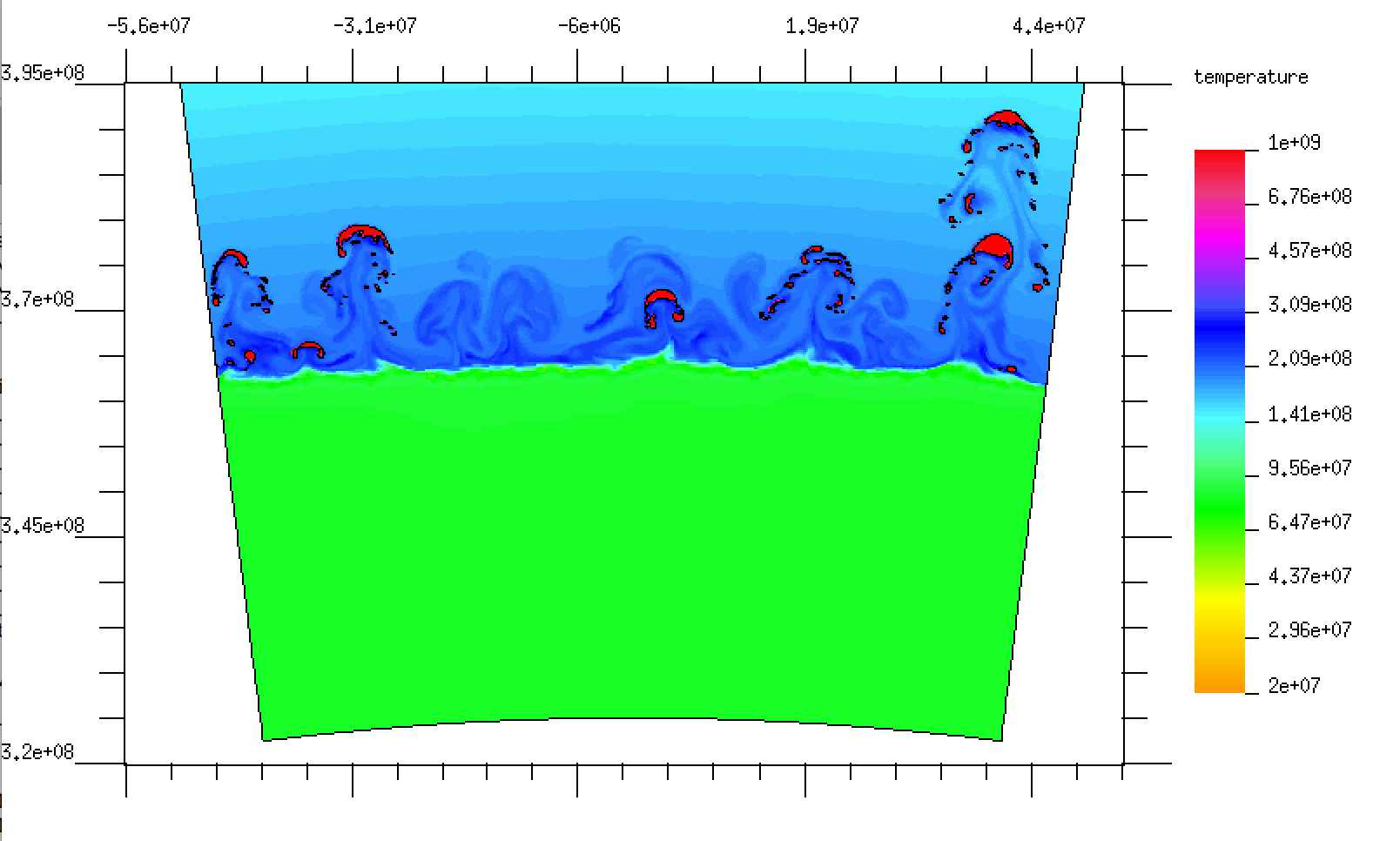}
	\caption{Temperature colour map at time 10.6s. Black contours define the limit between the two phases (see ~\ref{sec:vul}) }
	\label{fig:tmp_10.6}
\end{figure}

\begin{figure}
	
	\includegraphics[width=\columnwidth]{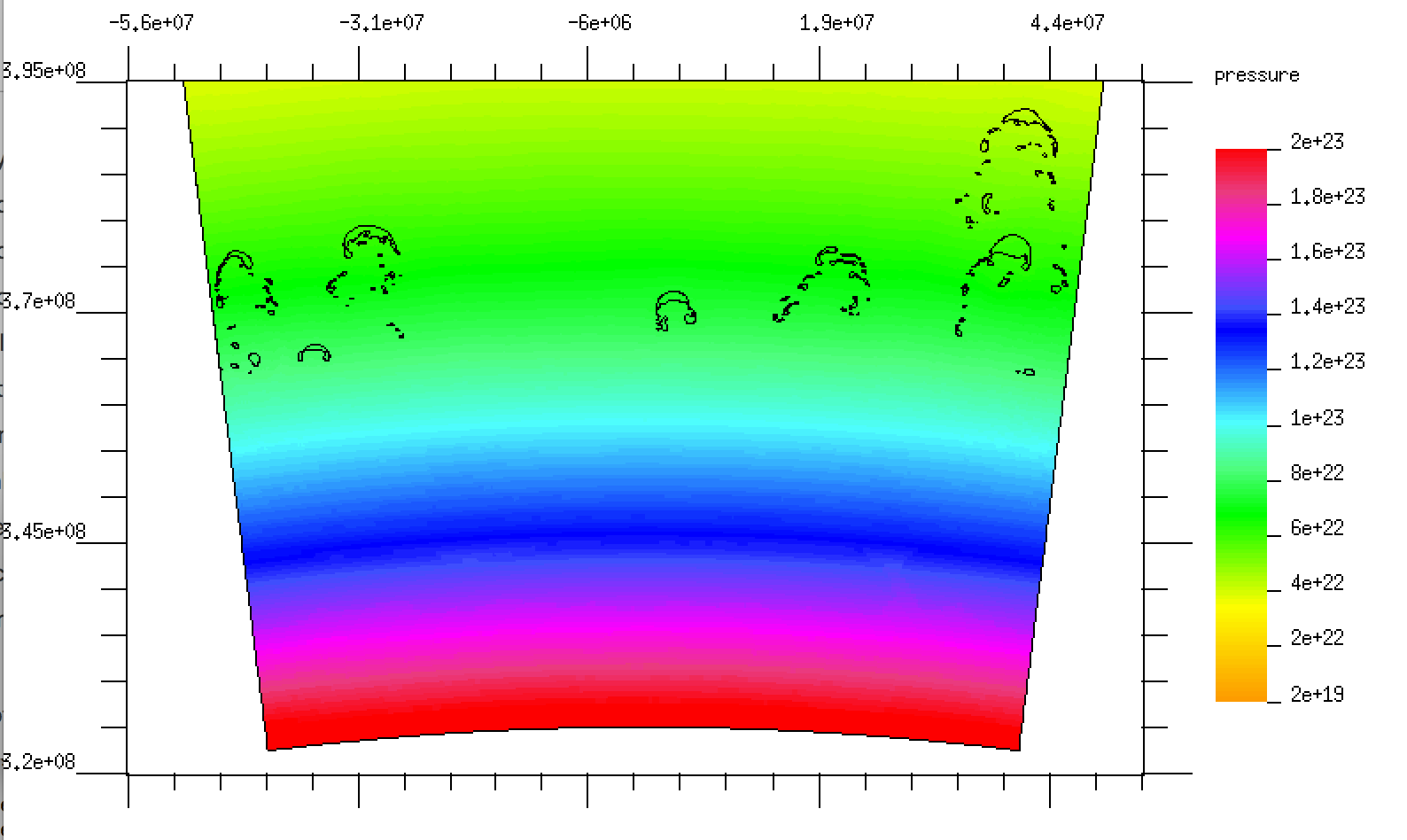}
	\caption{Pressure color map at time 10.6 seconds (\textsc{VULCAN} model A). Black contours define the limit between the two phases (see ~\ref{sec:vul})}
	\label{fig:p_10.6}
\end{figure}

      Since the configurations are unstable to convection, a small temperature
 perturbation develops very quickly (within $\approx 8$s) into a fully developed convective flow. For both solvers (\textsc{vulcan} and \textsc{rich}), the local convective burning heats small regions of matter to high temperatures, $\approx2.0\times 10^{9} \ K$. The small hotspots float as plumes in the envelope without any significant change in the pressure profile (Figs.~\ref{fig:tmp_10.6} and ~\ref{fig:p_10.6}).

\begin{figure}
	\includegraphics[width=\columnwidth]{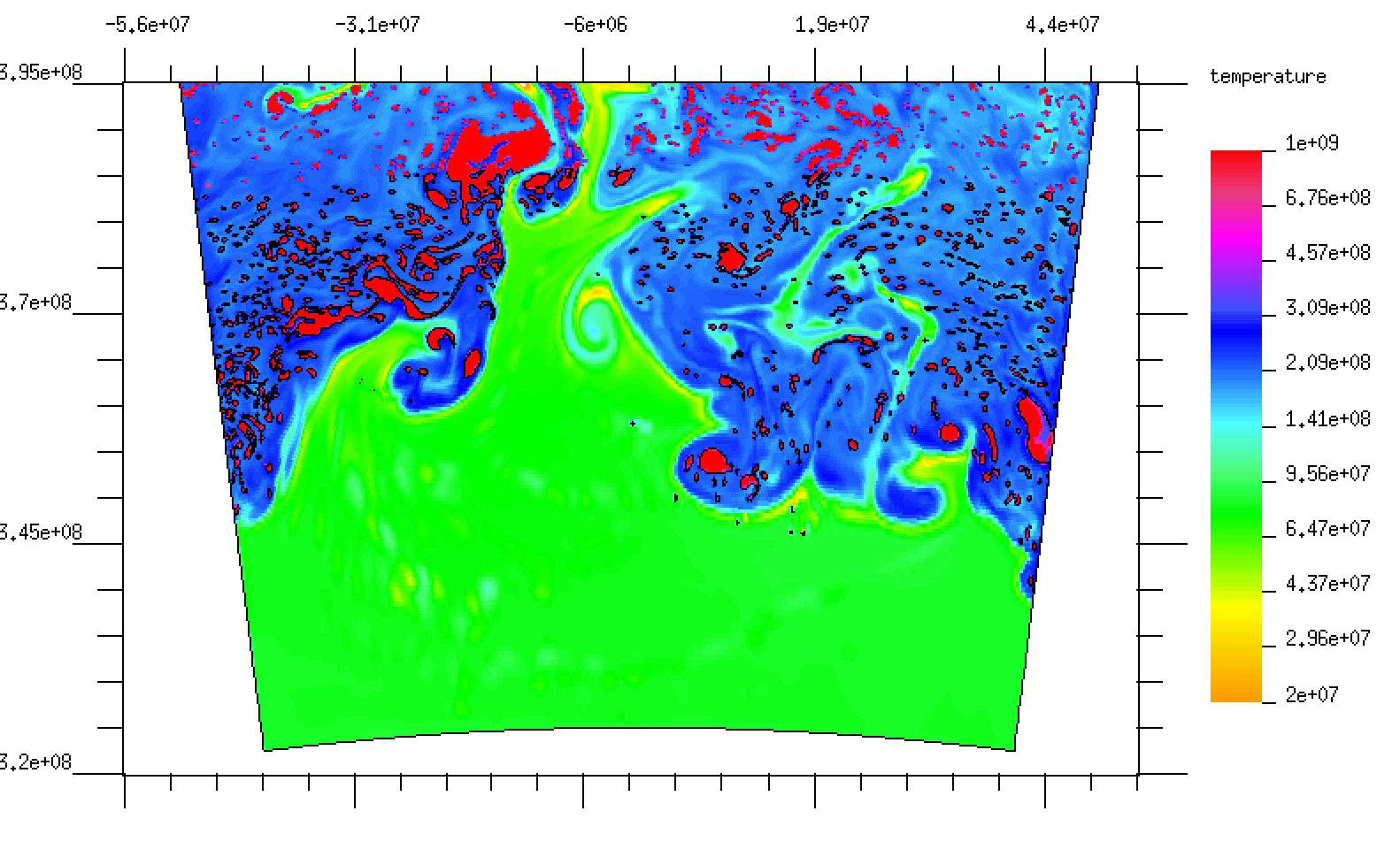}
	\caption{Temperature colour map at time 16.335 s. Black contours define the limit between the two phases (see ~\ref{sec:vul})}
	\label{fig:t_16.335}
\end{figure}

\begin{figure}	
	\includegraphics[width=\columnwidth]{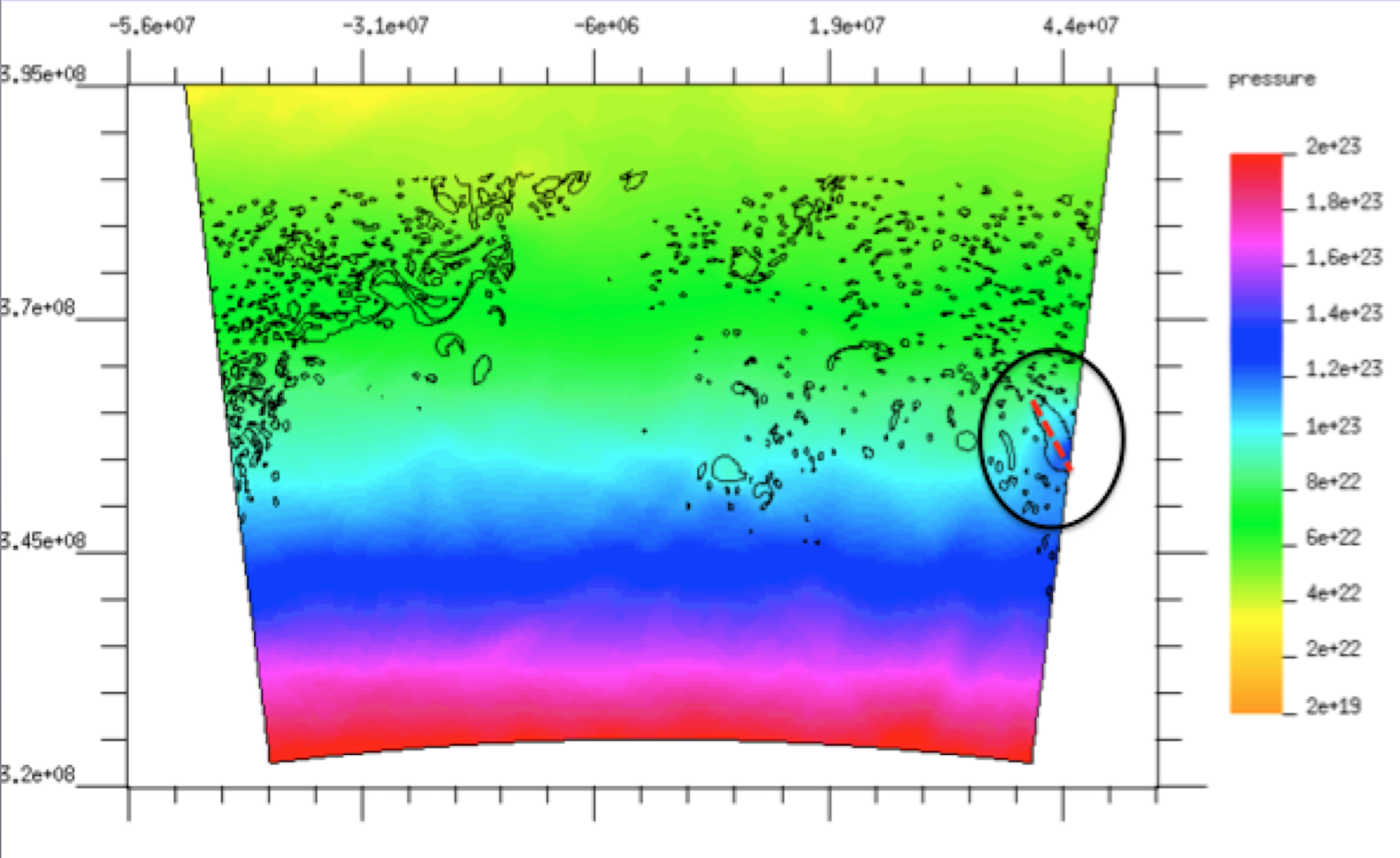}
	\caption{Pressure colour map at time 16.335 s (\textsc{vulcan} model A). Black contours define the limit between the two phases (see ~\ref{sec:vul})}
	\label{fig:p_16.335}
\end{figure}

As in the 1D model, heating by helium burning increases the background temperature and enhances the convective instability. Shear flow at the interface between the helium envelope and the CO core induces mixing and penetration of hot helium to the core. Up to time $t=~16.3$ s for the \textsc{vlucan} model and time $t=~18.3$ s for the \textsc{rich} model, the number and size of small plumes increase but they all float without any significant influence on the pressure profile. During this time, the interface between the CO of the core and the helium of the accreted envelope is distorted and hot helium penetrates deeper and deeper into the core. (Figs.~\ref{fig:t_16.335} and ~\ref{fig:p_16.335}).

\begin{figure}
 	\includegraphics[width=\columnwidth]{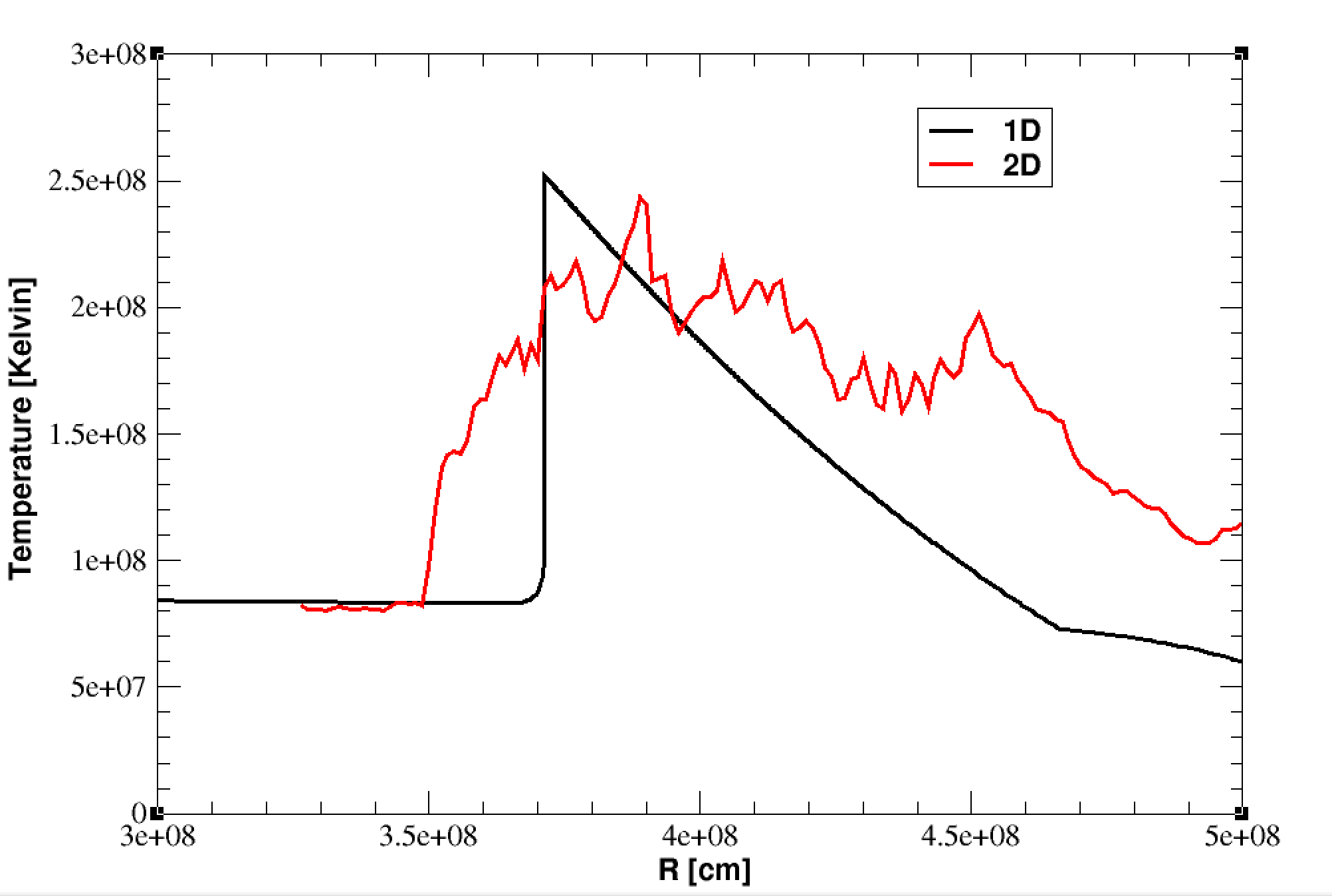}
 	\caption{The average 2D temperature profile - red versus the 1D temperature profile - black, at time 16.335 s.}
 	\label{fig:t_r_t16.335}
 \end{figure}

     Examining the transversally averaged radial temperature profile in the \textsc{vulcan} Fig.~\ref{fig:t_r_t16.335}, we notice two significant facts:
     \begin{enumerate}[label=(\alph*)]
\item In the 2D model the radial decrease of temperature profile is much shallower than in the 1D model.       
       \item The maximal average temperature reaches a maximum of $\approx2.5\times 10^{8} \ K$, and in Fig.~\ref{fig:tau} we see that indeed its the first time that the time-scale hierarchy changes and $\tau_{burn} \le \tau_{hyd}$.
       \end{enumerate}

\begin{figure}
	
	\includegraphics[width=\columnwidth]{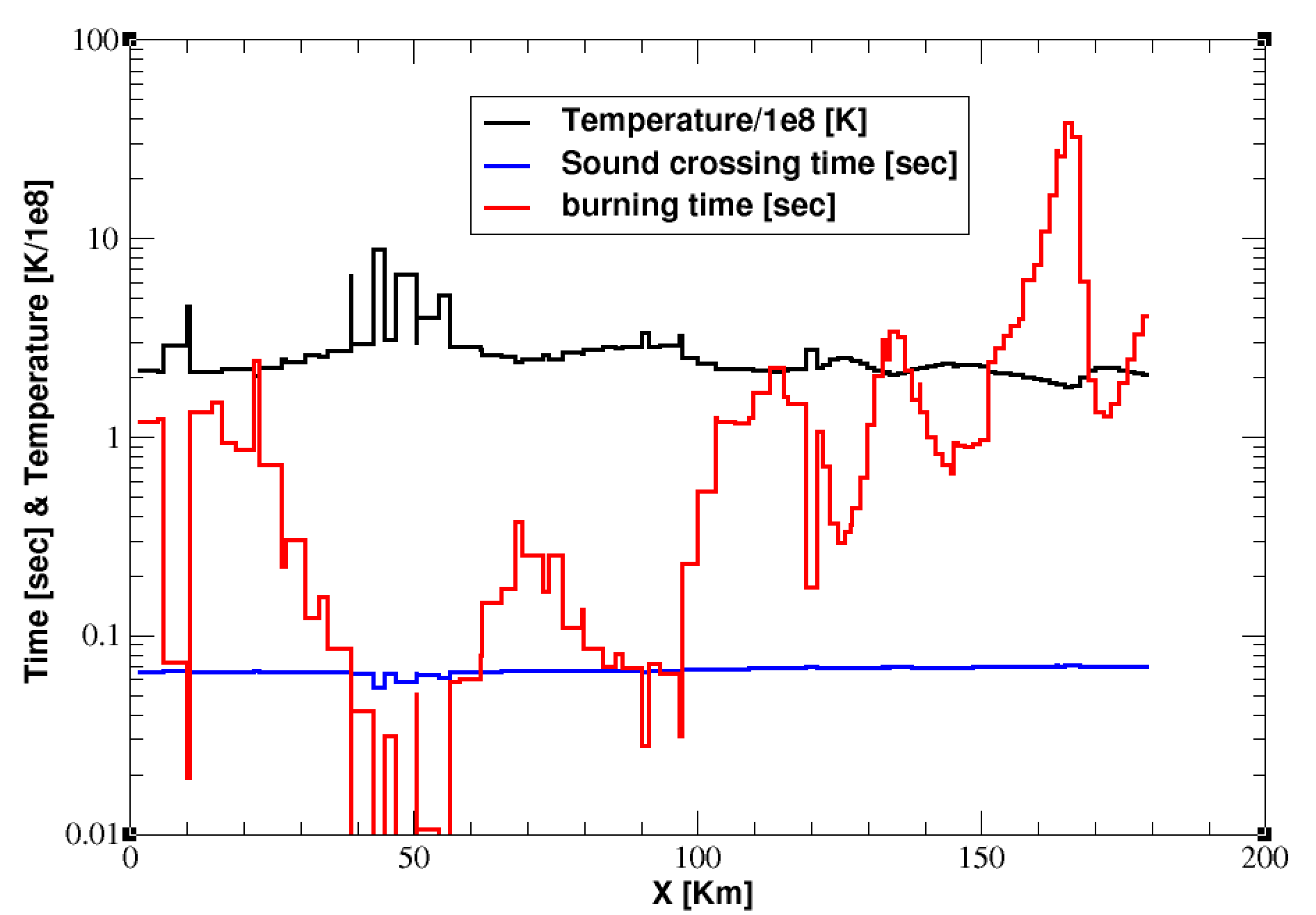}
	\caption{The temperature (black) and relevant time-scales profile (red - $\tau_{burn} $, blue - $\tau_{hyd}$) along a line in the ignition region at time 16.335 s (see text and the marked region in Fig.~\ref{fig:p_16.335})}
	\label{fig:times_16.335}
\end{figure}

      Examining the pressure distribution at this stage (Fig.~\ref{fig:p_16.335}), we see (marked region at the right side of the figure) a region  of helium, with dimensions of a few kilometres, that penetrates deep into the core. For this region, the temperature increase leads to the development of local high pressure. 
     The enhanced burning develops within $ \approx0.04$ s into a detonation. In order to study this ignition stage we extracted temperature and density profiles along a line in this local region (marked in dashed red line in Fig.~\ref{fig:p_16.335}) and calculated $\tau_{burn}$ and $\tau_{hyd}$ along this line. In Fig.~\ref{fig:times_16.335}, we indeed notice that in this marked zone in a region of $\approx30$ kilometres the local burning time is much shorter than the local hydrodynamic time.
     
\begin{figure}	
	\includegraphics[width=\columnwidth]{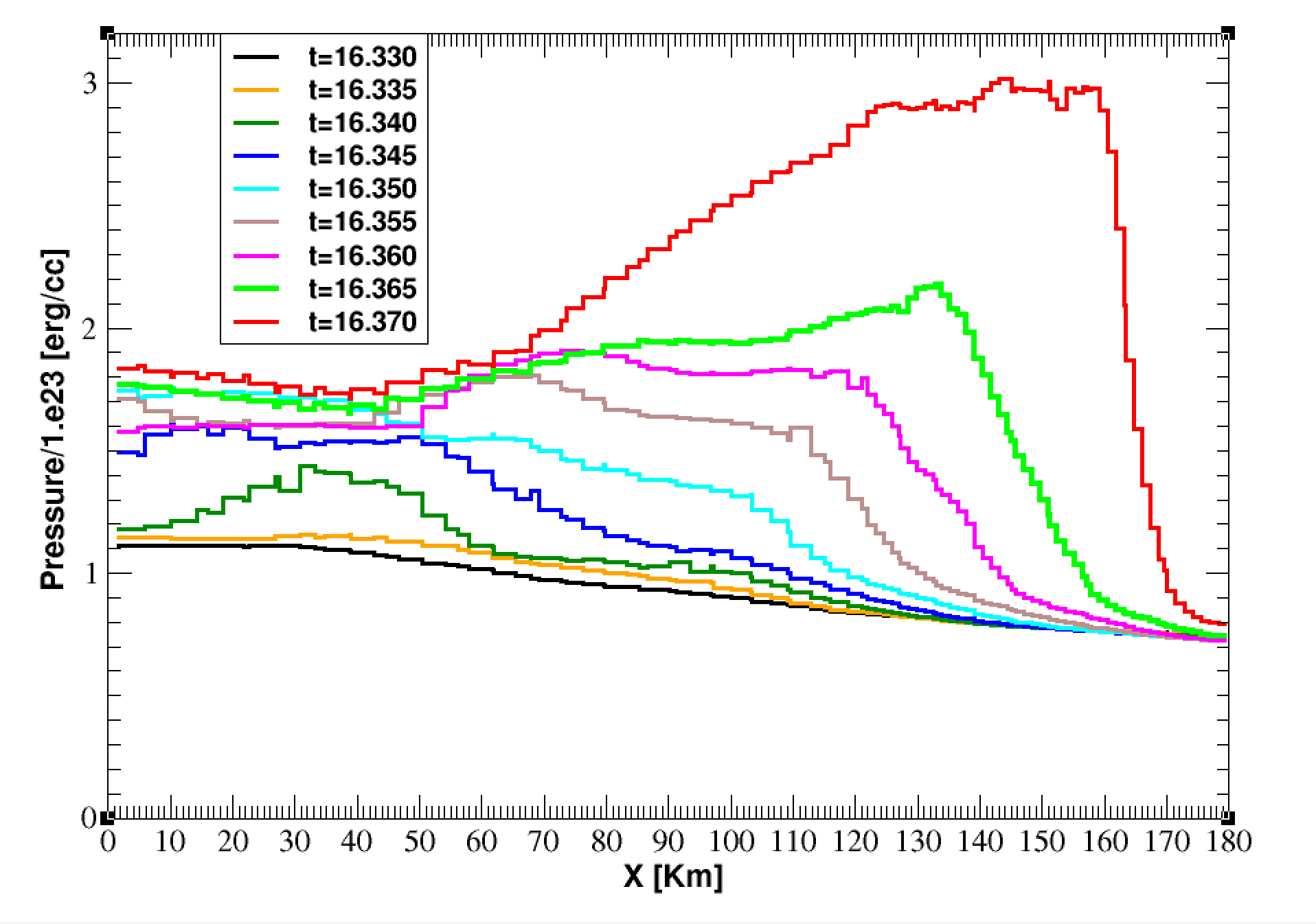}
	\caption{Pressure profiles along a line in the ignited region (Fig.~\ref{fig:p_16.335}). The first profile (black) is at time 16.330 seconds and the last one (red) is at time 16.370 s (\textsc{vulcan} model A).}
	\label{fig:PH2c_1_9_cut_p}
\end{figure}

      The stages leading to detonation are clearly observed along the line marked in Fig.~\ref{fig:p_16.335} for which we present in Fig.\ref{fig:PH2c_1_9_cut_p}  a time series of pressure profiles for. In the first profile (black) at time t=16.330 s the pressure is smooth and in perfect match with the average radial pressure. At time t=16.340 s (dark green) we see the first signs of increase in the pressure. The elevated pressure occurs in a zone that is consistent with the time-scales presented in  Fig.\ref{fig:times_16.335}.  

\begin{figure}
 	\centering
 	\includegraphics[width=0.9\linewidth]{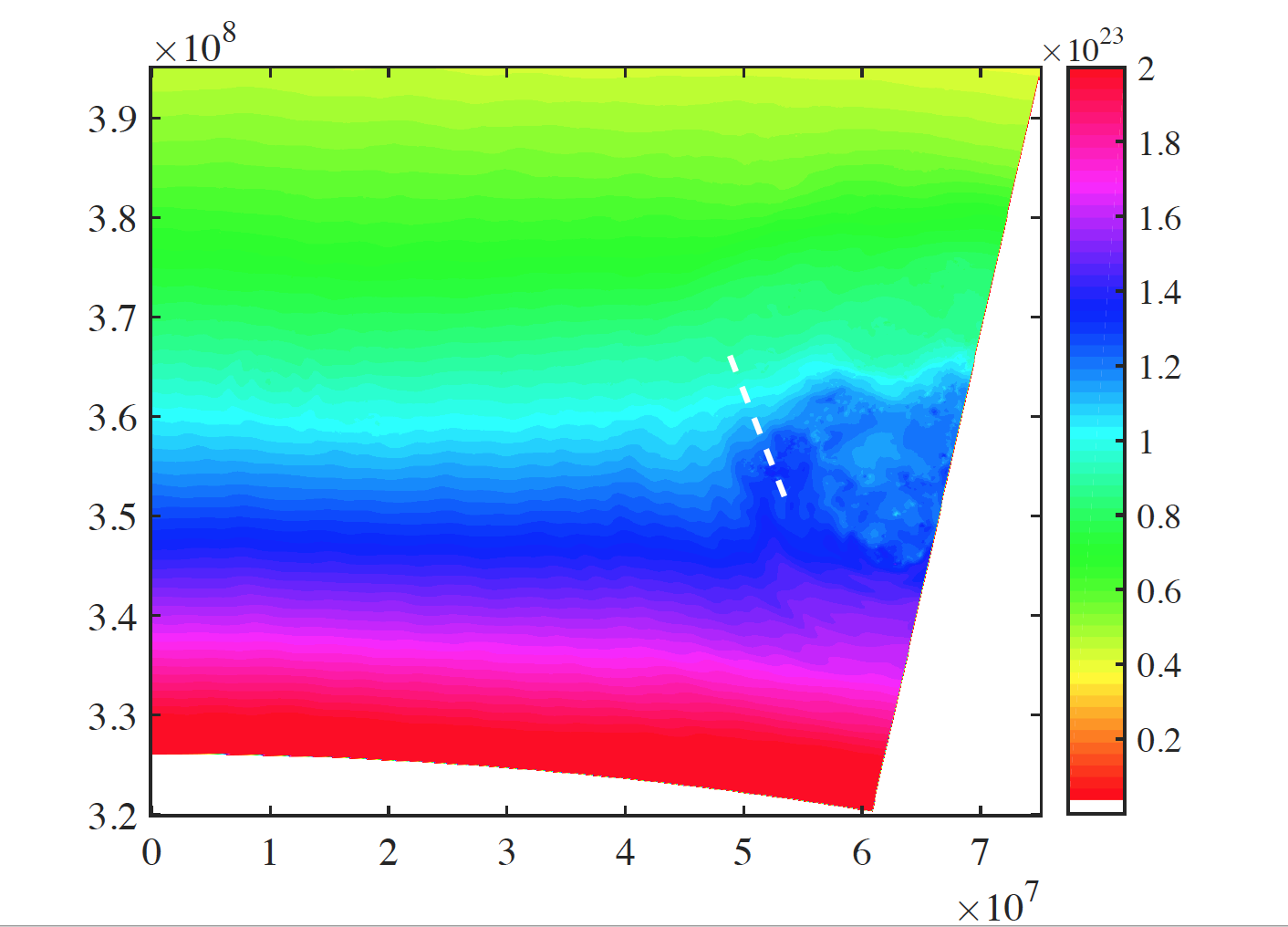}
 	\caption{The pressure distribution at time $t=18.366$ s (model A \textsc{rich}). The dashed white line is the curve along which the data is extracted in Figs \ref{fig:times_18_366} and \ref{fig:rich_cut_p}.}
 	\label{fig:p_18_366}
 \end{figure}

\begin{figure}
 	\centering
 	\includegraphics[width=0.9\linewidth]{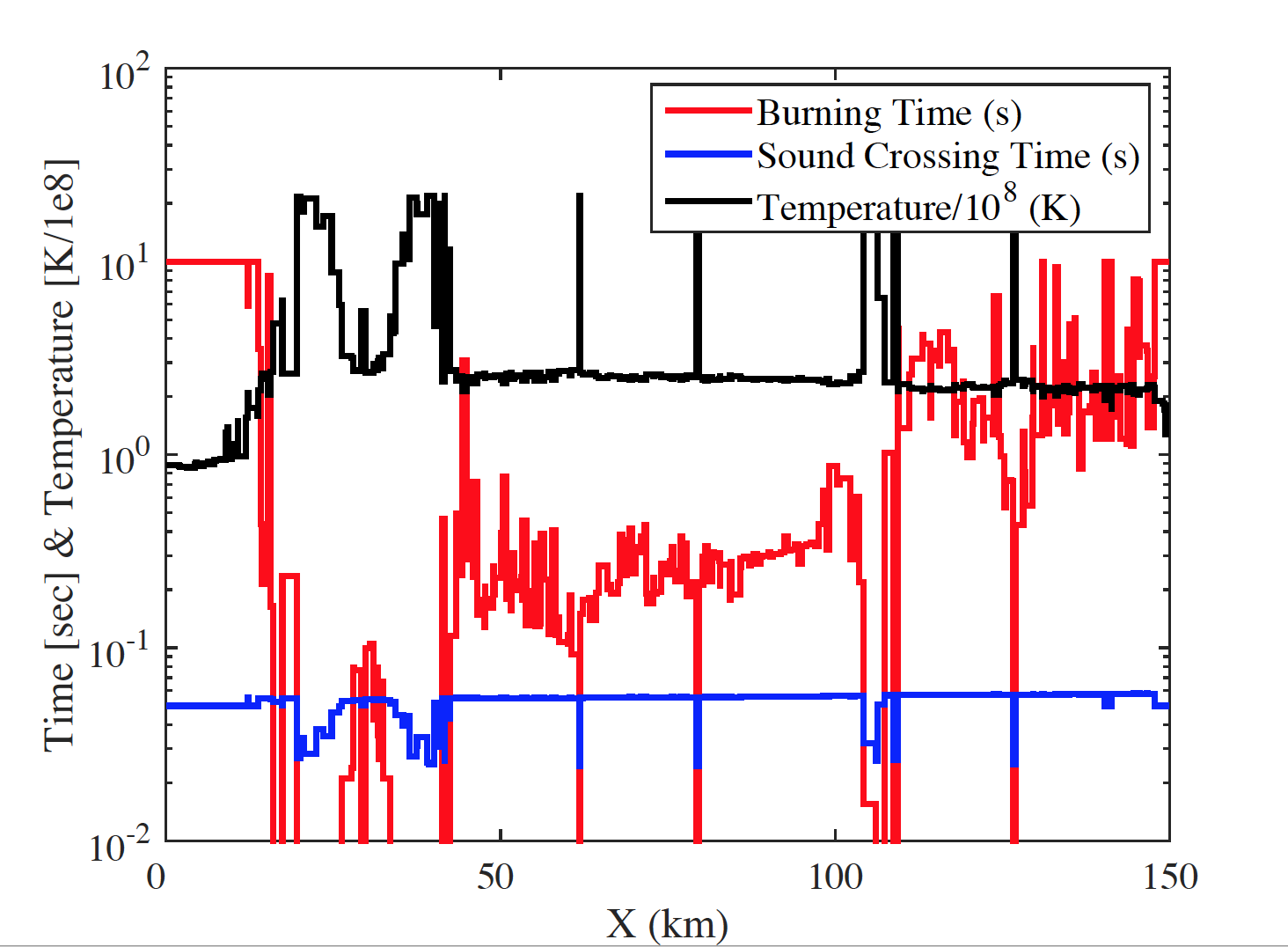}

 	\caption{The temperature (black) and relevant time-scales profile (red - $\tau_{burn} $, blue - $\tau_{hyd}$) along a line in the ignition region at time $t=18.366$ s along the marked line shown in fig. \ref{fig:p_18_366}(\textsc{rich} model A).}
 	\label{fig:times_18_366}
 \end{figure}

\begin{figure}
 	\centering
 	\includegraphics[width=0.9\linewidth]{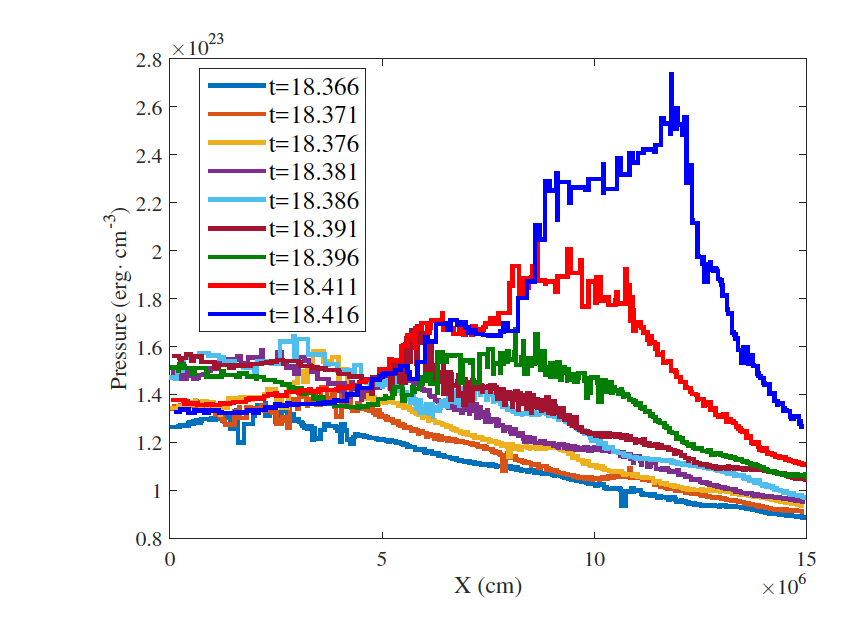}
 	\caption{The pressure profile along the marked line in fig. \ref{fig:p_18_366} at various times (\textsc{RICH} model A).}
 	\label{fig:rich_cut_p}
 \end{figure}

      It is encouraging to find that in the \textsc{rich} run the sequence of events is almost exactly the same as in the \textsc{vulcan}. 
      In Fig.~\ref{fig:times_18_366}, we indeed notice that for the marked line in the pressure colour map at time t=18.366 s   (Fig.~\ref{fig:p_18_366}) there is a region of $\approx30$ kilometres for which the local burning time is much shorter than the local hydrodynamic time. 
      The time series of pressure profiles ()Fig.\ref{fig:rich_cut_p}) along the line marked in Fig.~\ref{fig:p_18_366} are in perfect agreement with the \textsc{vulcan} results.
      
\begin{figure}
	
	\includegraphics[width=\columnwidth]{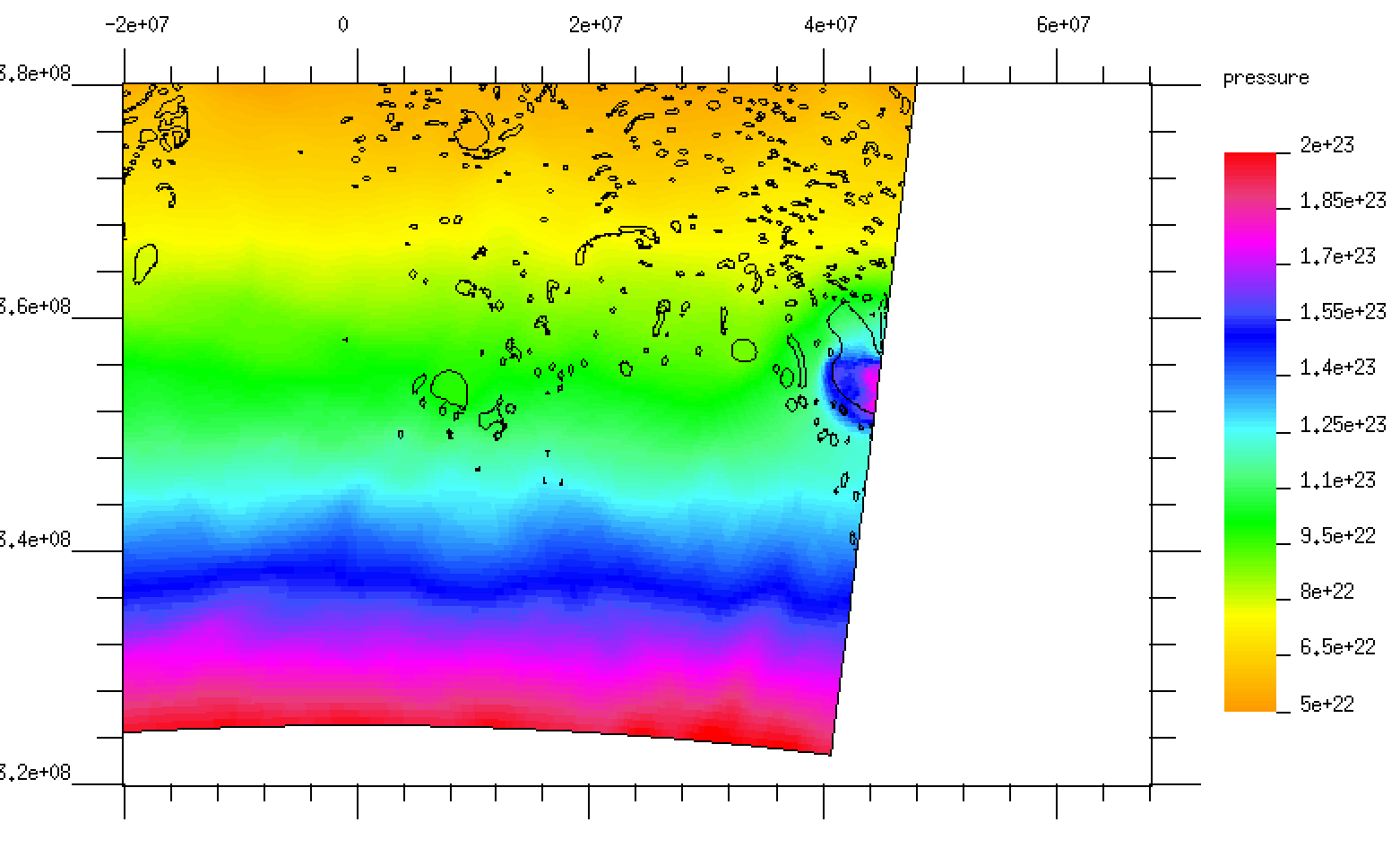}
	\caption{Pressure colour map of the full 2D region at time t=16.345 s (\textsc{vulcan} model A).}
	\label{fig:PH2c_t16_345_p}
\end{figure}

\begin{figure}
	
	\includegraphics[width=\columnwidth]{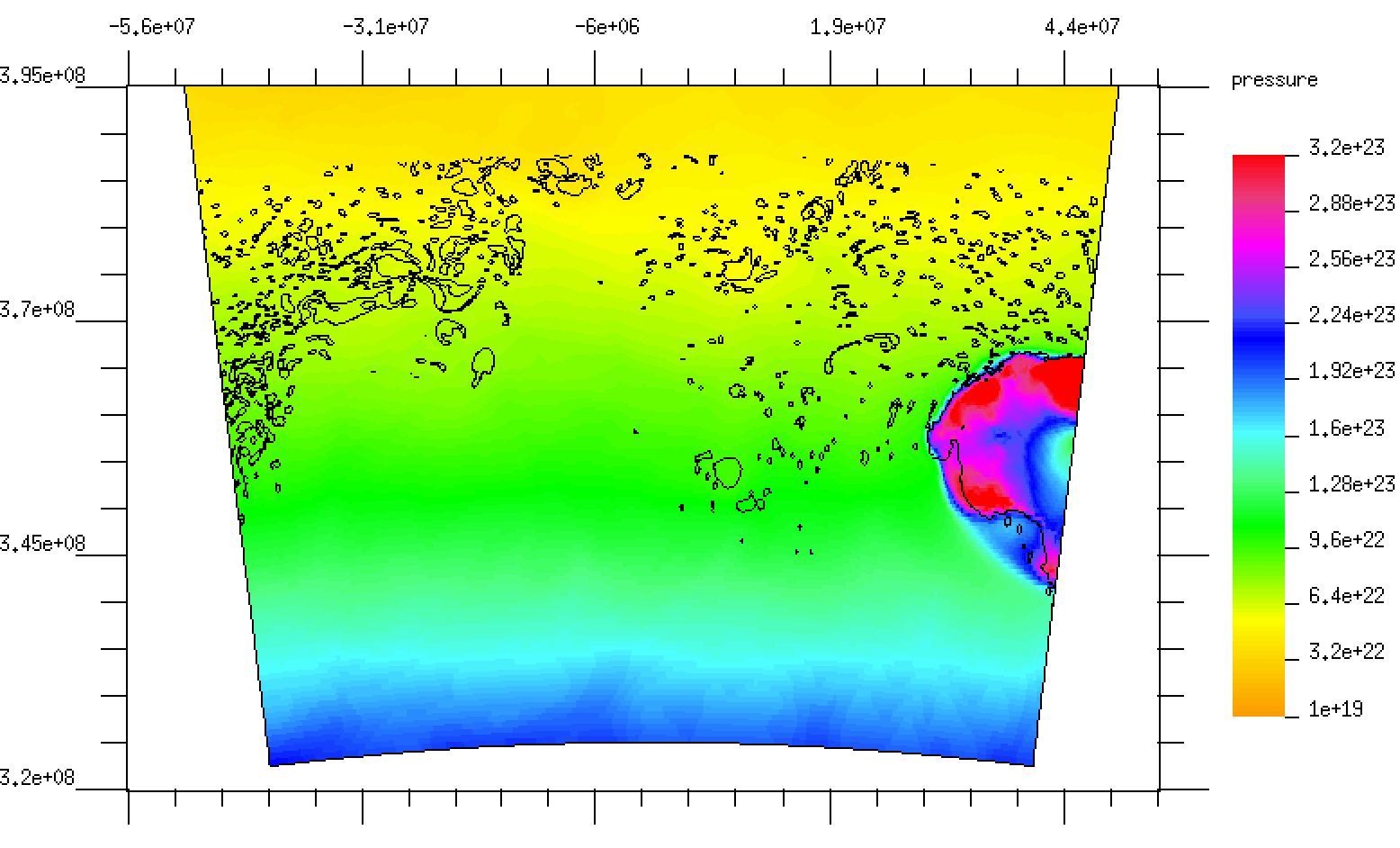}
	\caption{Pressure colour map of the full 2D region at time t=16.370 s (\textsc{vulcan} model A).}
	\label{fig:PH2c_t16_370_p}
\end{figure}

\begin{figure}
	
	\includegraphics[width=\columnwidth]{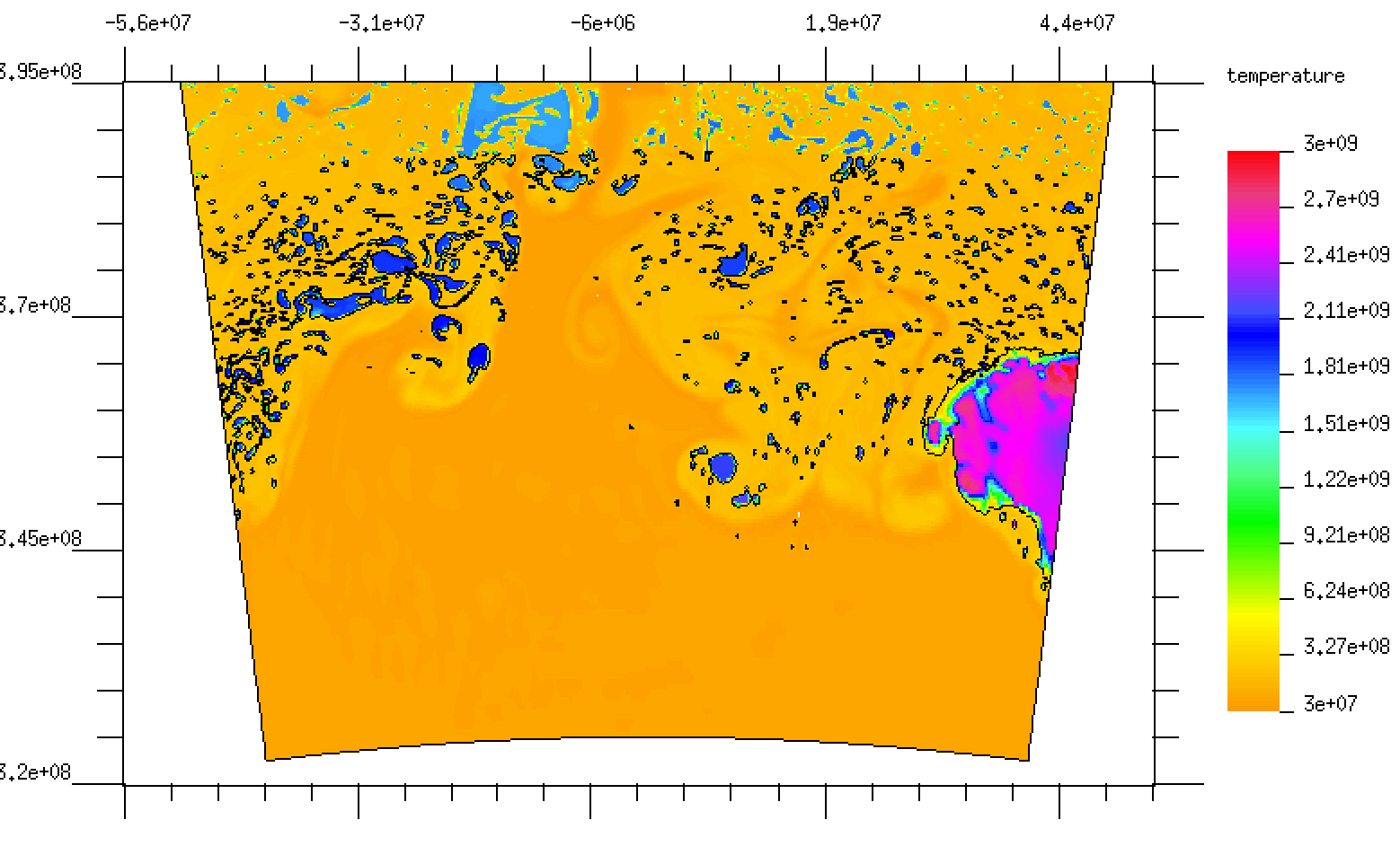}
	\caption{Temperature colour map of the full 2D region at time t=16.370 s (\textsc{vulcan} model A).}
	\label{fig:PH2c_t16_370_T}
\end{figure}

      For the \textsc{vulcan} model at time t=16.370 s, the pressure increases by a factor of 3 in a region with dimensions of about a hundred kilometres and the flow is already almost a fully developed detonation.
        The full 2D pressure colour map, at time t=16.345 s, is presented in Fig.\ref{fig:PH2c_t16_345_p}  and at time 16.370 s in Fig.\ref{fig:PH2c_t16_370_p}. The full 2D temperature colour map, at time t=16.370 s, is presented in Fig.\ref{fig:PH2c_t16_370_T}.

\begin{figure}
\centering
\includegraphics[width=0.9\linewidth]{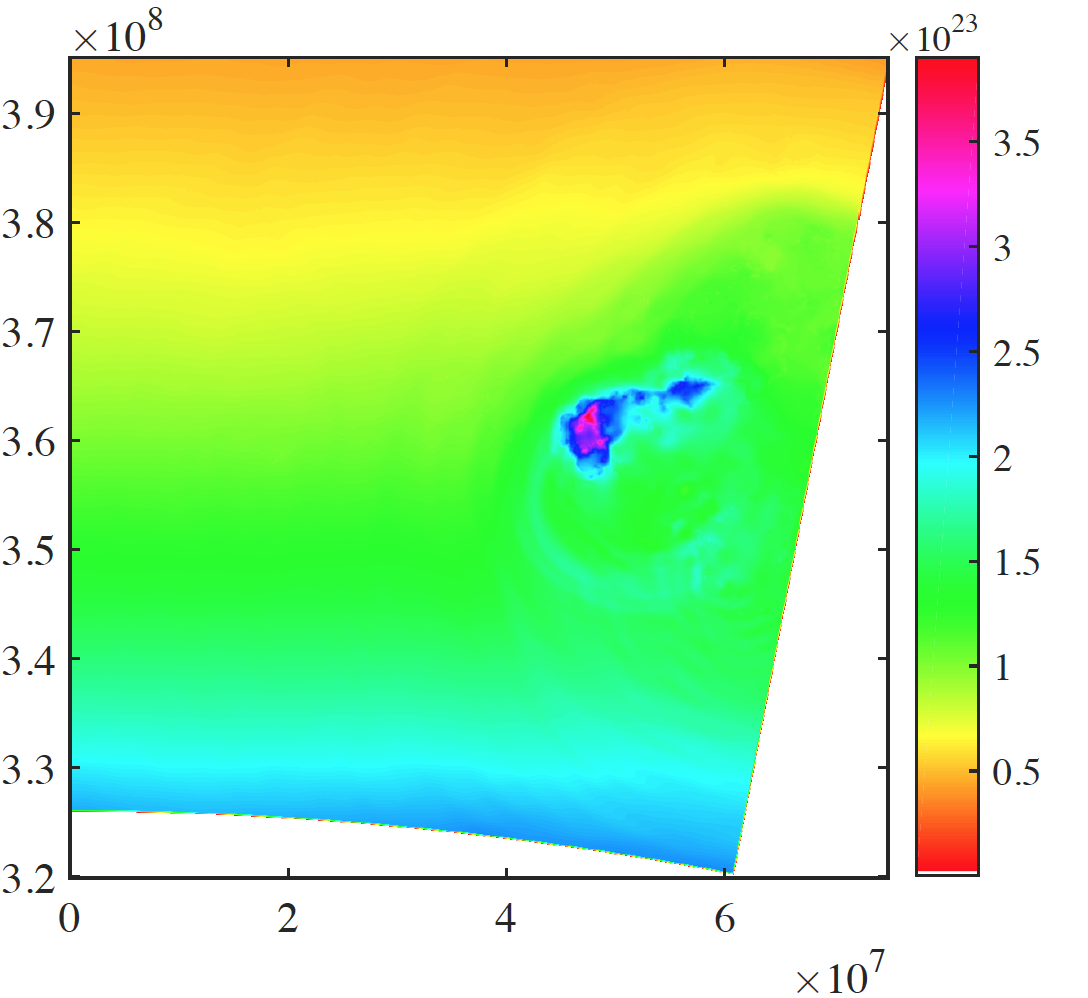}
\caption{The pressure distribution at time $t=18.416$ s showing the initial propagation of the detonation (\textsc{rich} model A).}
\label{fig:pressure_end}
\end{figure}

\begin{figure}
	\centering
	\includegraphics[width=0.9\linewidth]{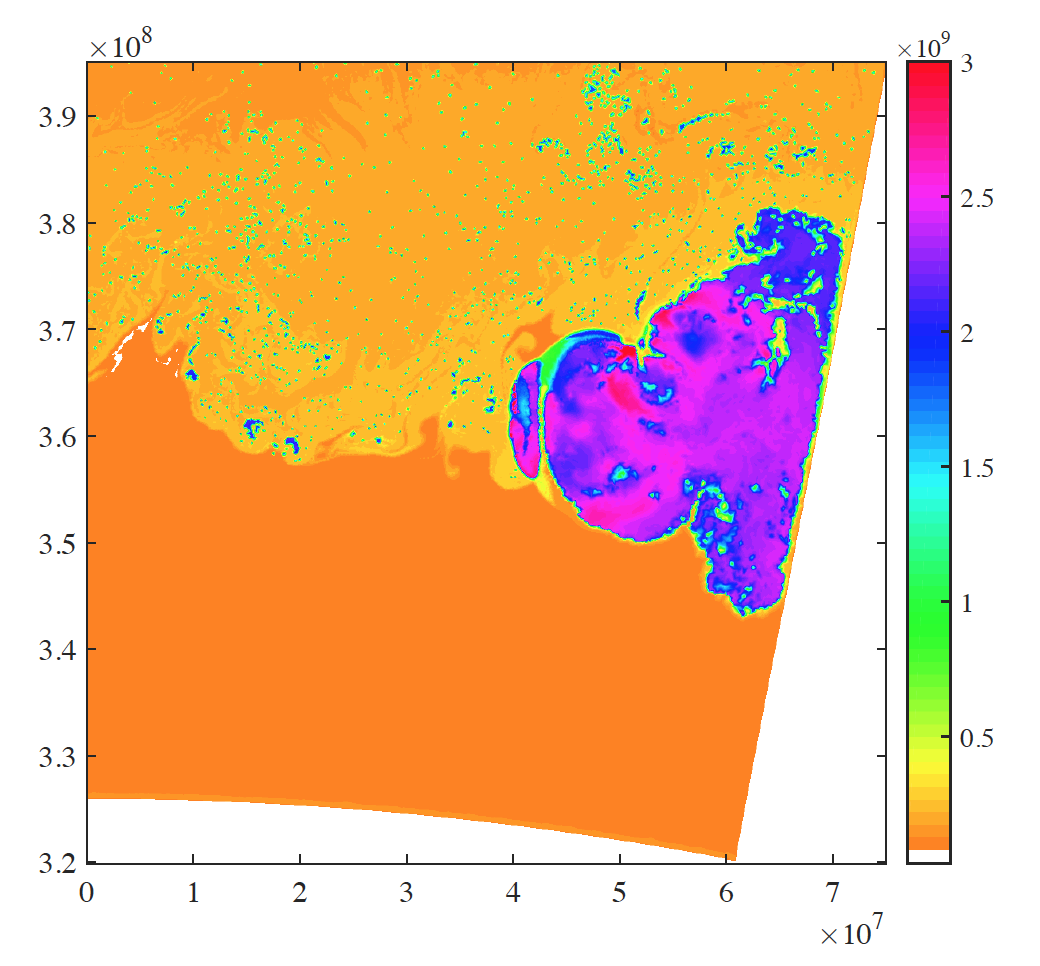}
	\caption{The temperature distribution at time $t=18.416$ s (\textsc{rich} model A).}
\label{fig:temperature_end}
\end{figure}

     The \textsc{rich} model shows exactly the same features.  The full 2D pressure colour map, at time t=18.416 s, is presented in Fig.\ref{fig:pressure_end}. The full 2D temperature colour map, at time t=18.416 s, is presented in Fig.\ref{fig:temperature_end}.

\begin{figure}	
  	\includegraphics[width=\columnwidth]{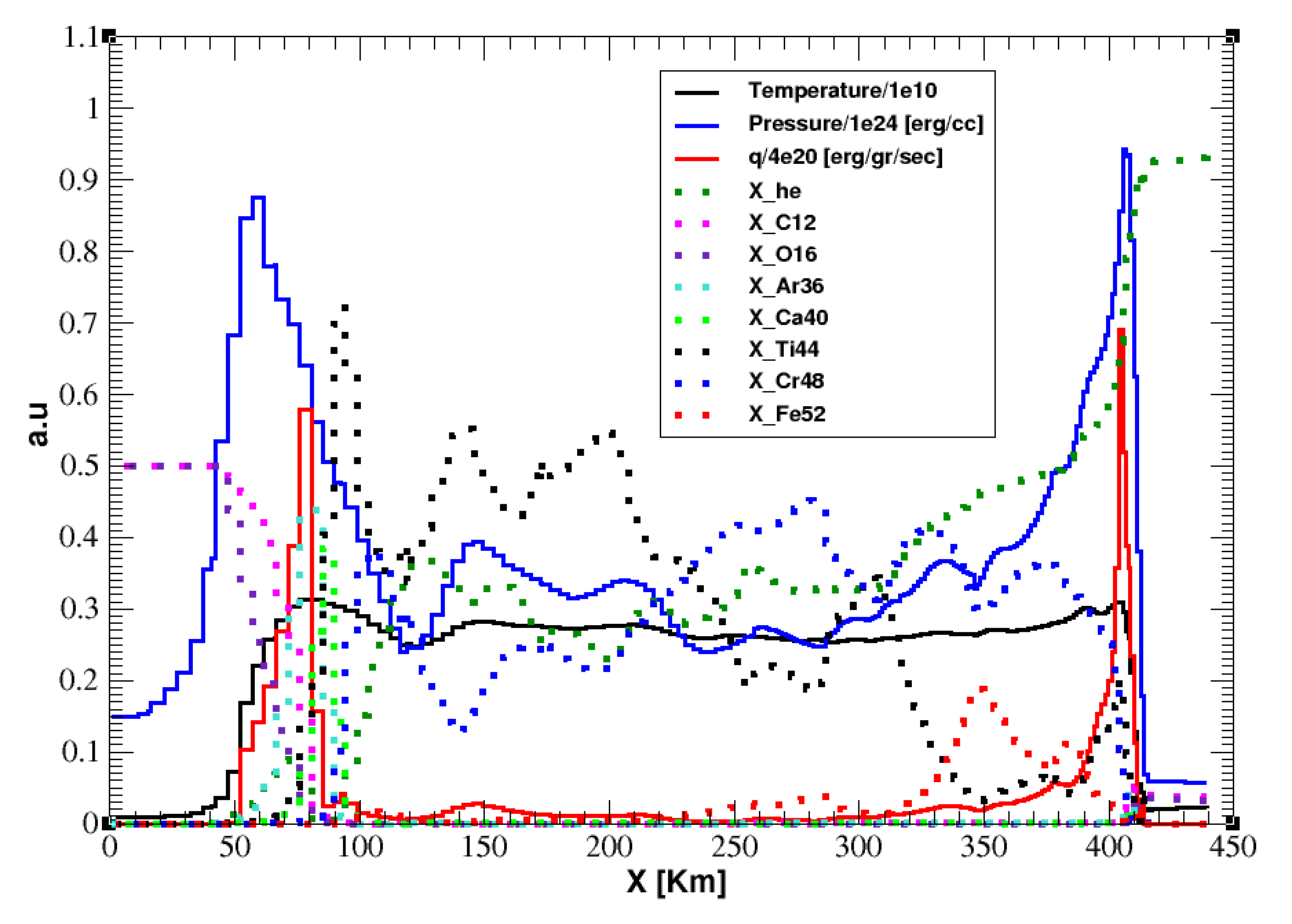}
  	\caption{Pressure (blue), temperature (black), reaction rates (red) and abundance (dotted) profiles along a line in the detonated region. The spatial coordinate is from inside (CO rich) to the outer regions (He rich) at time 16.385 s (\textsc{vulcan} model A).}
  	\label{fig:HEL_1m_PH2e_t16_385_cut}
  \end{figure}

  The fully developed detonation propagates outwards into the helium-rich outer envelope and inwards into the CO core material. In Fig.~\ref{fig:HEL_1m_PH2e_t16_385_cut}, we present the pressure, temperature, reaction rates, and abundance profiles along a line in the detonated region.
   At the left-hand side, facing the CO core, we observe the detonation front that burns the CO mostly into intermediate mass elements, mainly \iso{44}Ti and \iso{48}Cr. At the right-hand side, we see the outgoing detonation advancing into the helium envelope. The burning products here are also mainly intermediate mass elements with a few percent of iron group elements. 
  Although the computational domain is a bit too small it seems that for this model we get a double detonation and the detonation indeed penetrates into the CO core.
  
\begin{figure}
  	
  	\includegraphics[width=\columnwidth]{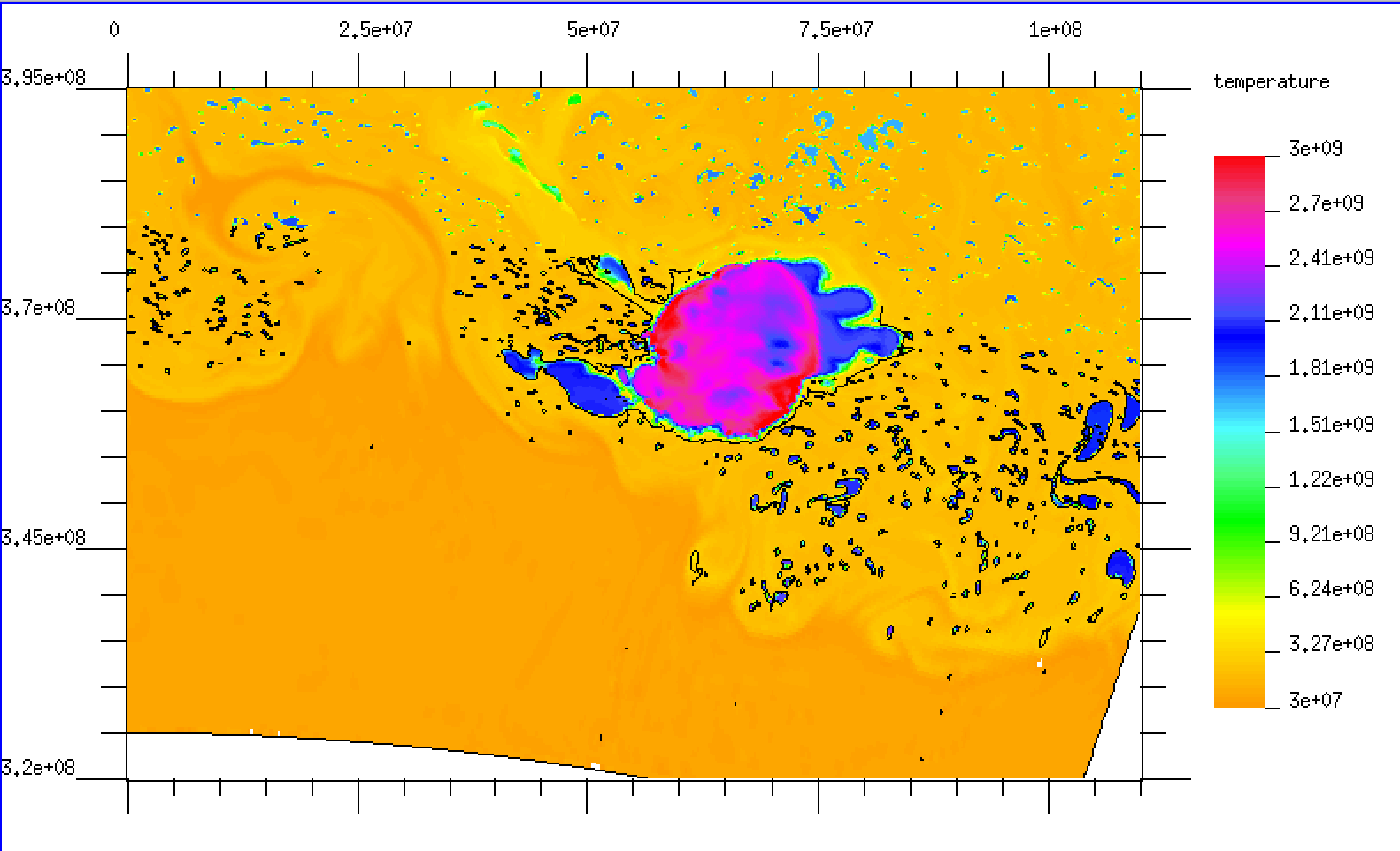}
  	\caption{Temperature colour map of the full 2D region at time t=15.596 s (\textsc{vulcan} model AW).}
  	\label{fig:HEL_1m_PH2g_T}
  \end{figure}

\begin{figure}
	
	\includegraphics[width=\columnwidth]{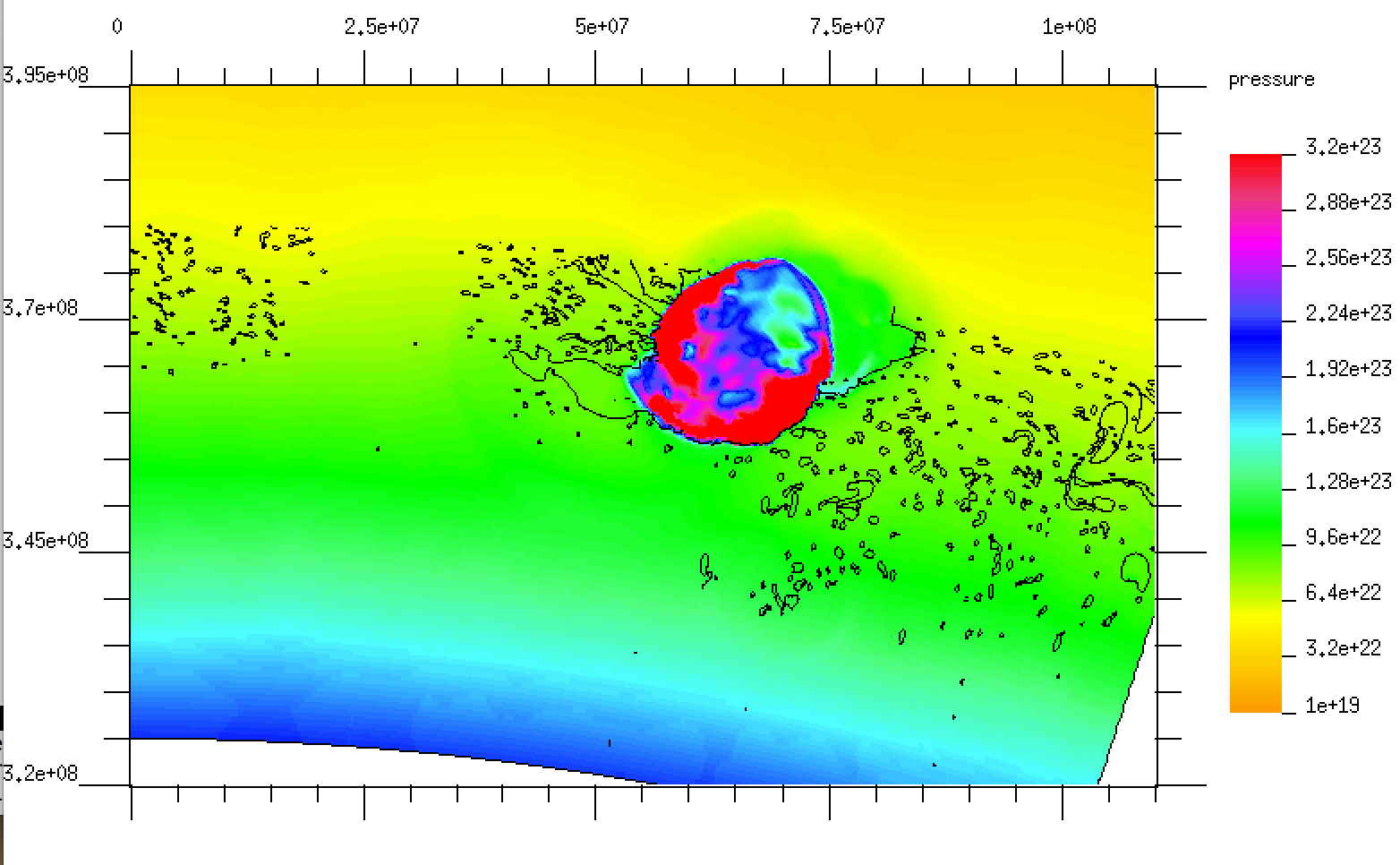}
	\caption{Pressure color map of the full 2D region at time t=15.596 seconds (\textsc{VULCAN} model AW).}
	\label{fig:HEL_1m_PH2g_P}
\end{figure}

  The near edge ignition of model A is not a characteristic feature; in many other cases the ignition occurs in the central regions as presented in model AW (Figs.\ref{fig:HEL_1m_PH2g_T} and \ref{fig:HEL_1m_PH2g_P}; see also model BW).
  
\begin{figure}
	
	\includegraphics[width=\columnwidth]{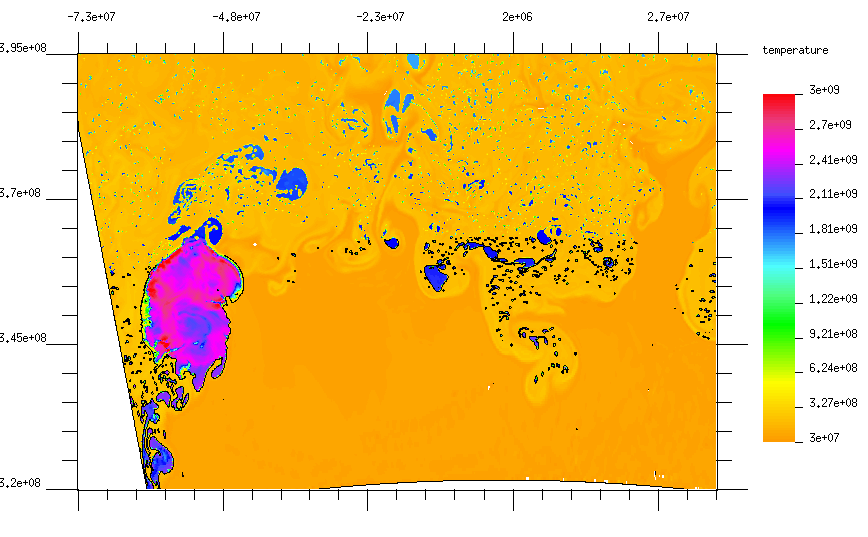}
	\caption{Temperature colour map of the full 2D region at time t=10.50 s (\textsc{vulcan} model AD).}
	\label{fig:HEL_1m_PH2_9ee_t10_50_T}
\end{figure}

\begin{figure}
	
	\includegraphics[width=\columnwidth]{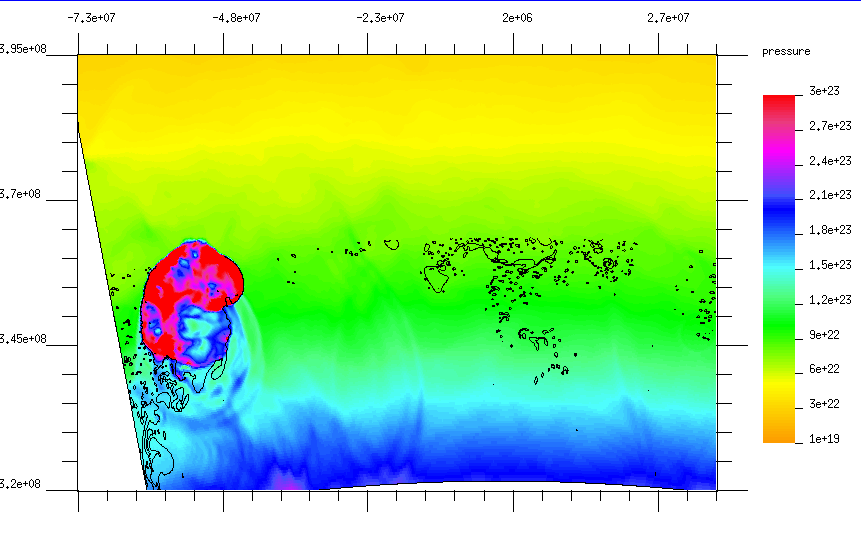}
	\caption{Pressure colour map of the full 2D region at time t=10.50 s (\textsc{vulcan} model AD).}
	\label{fig:HEL_1m_PH2_9ee_t10_50_P}
\end{figure}

    Models with more delicate resolution such as model AD (Table~\ref{tab:models}) show the same features and ignite a helium detonation in the region that is mixed by convection between the CO core and accreted helium envelop (Figs.\ref{fig:HEL_1m_PH2_9ee_t10_50_T} and \ref{fig:HEL_1m_PH2_9ee_t10_50_P}). The more delicate model is fully consistent with all the A type models presented here.

\clearpage

\subsection{The evolution to ignition of models of type B}

\begin{figure}	
	\includegraphics[width=\columnwidth]{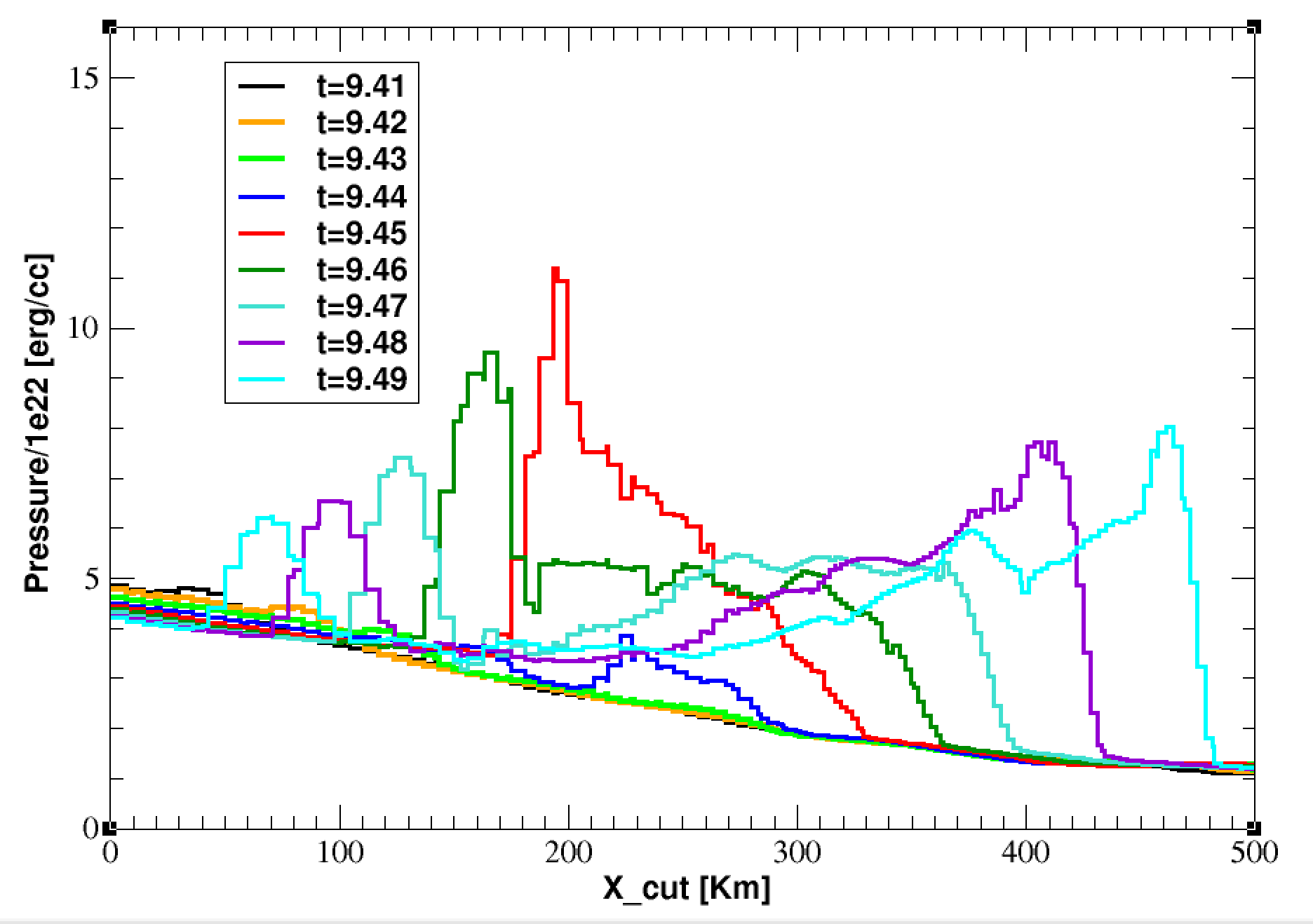}
	\caption{Pressure profiles along a line in the ignited region of model BW (\textsc{vulcan}). The inner coordinates face the CO core and the outer coordinates face the helium envelope. The first profile (black) is at time 9.41 s and the last one (cyan) is at time 9.49 s.}
	\label{fig:HEL_C1_E_05F_cut_p_all}
\end{figure}

\begin{figure}	
\includegraphics[width=\columnwidth]{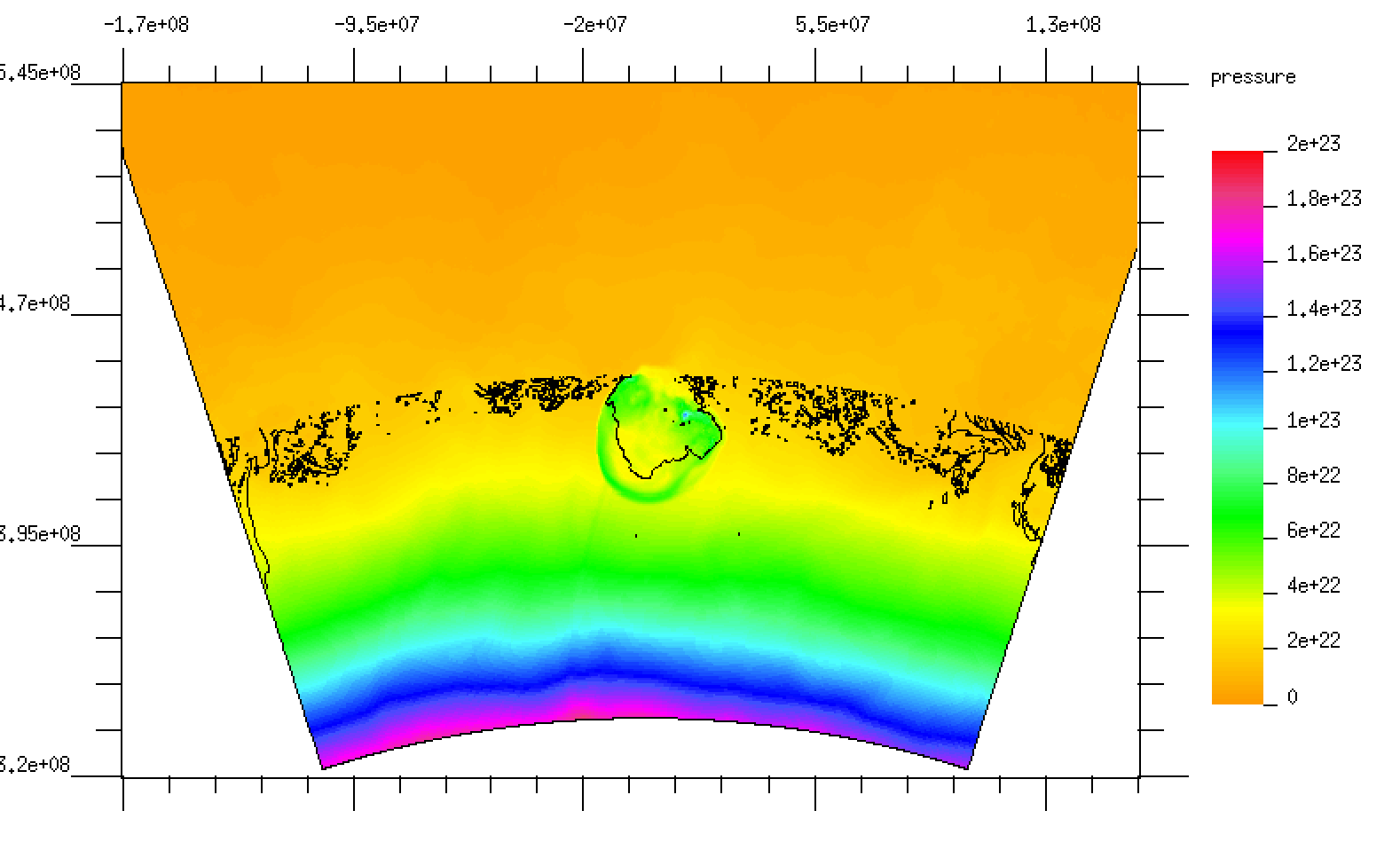}
\caption{Pressure colour map of the full 2D region of model BW (\textsc{vulcan}) at time t=9.49 s.}
\label{fig:HEL_C1_E_05F_cut_p_t9_49}
\end{figure}

\begin{figure}	
	\includegraphics[width=\columnwidth]{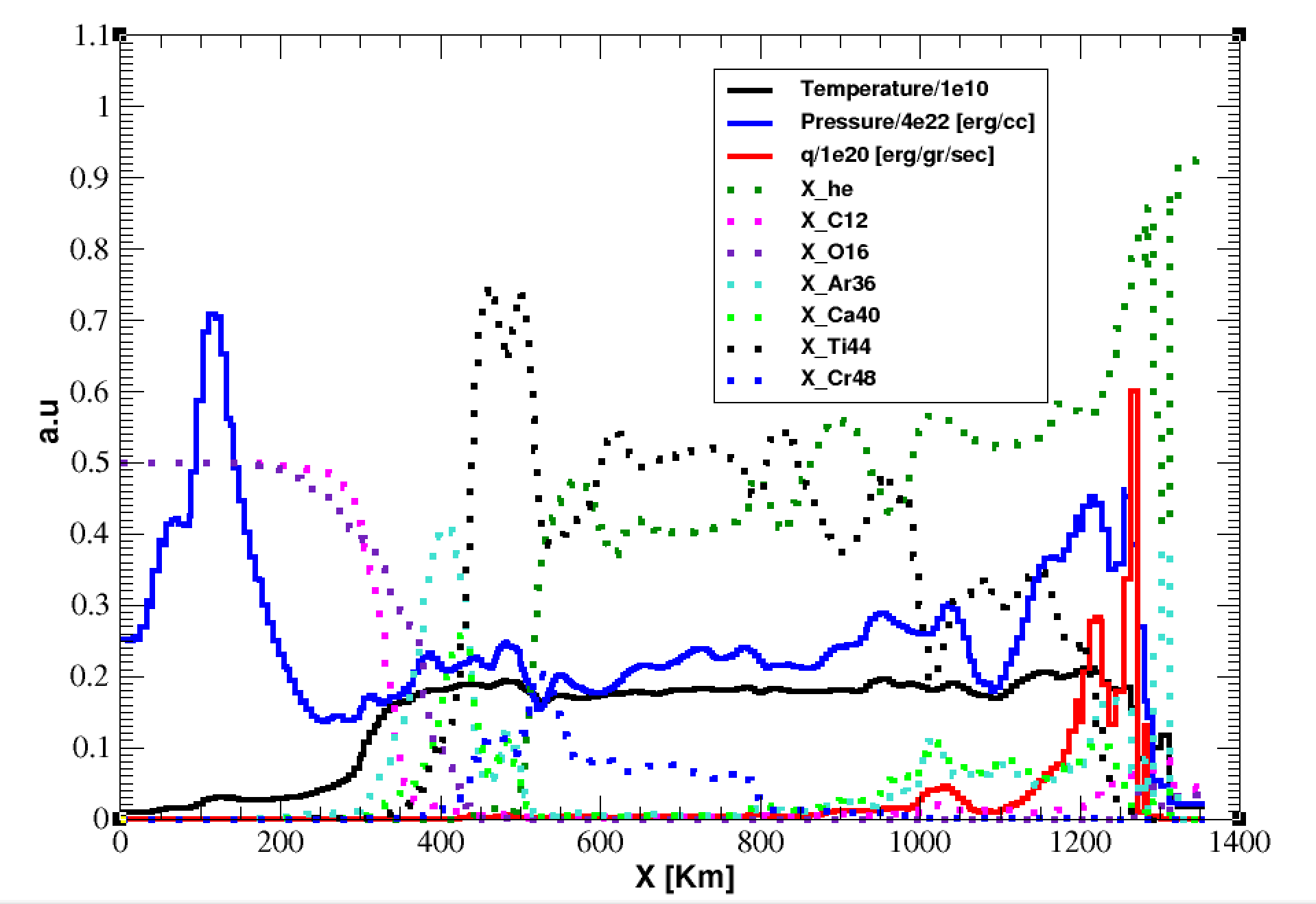}
	\caption{Pressure (blue), temperature (black), reaction rates (red) and abundance (dotted) profiles along a line in the detonated region. The spatial coordinate is from inside (CO rich) to the outer regions (He rich) at time 9.6 s (\textsc{vulcan} model BW).}
	\label{fig:HEL_C1_E_05F_t9_6_cut_short}
\end{figure}

  For models of type B the accretion rate is higher, the compression is more efficient and the heating of the envelope in the 1D model is faster. Convection sets in when the temperature at the base of the helium envelope is $\approx1.0\times 10^{8}  \ K$. In the 1D model, it takes a few years until the time-scales hierarchy changes and runaway conditions exist when the temperature is $\approx3.0\times 10^{8} \ K$. We started the 2D simulations when the base temperature was $\approx2.5\times 10^{8} \ K$ (in the 1D model it takes 20 s for the  temperature to raise from $2.4\times 10^{8} \ K$ to $3.0\times 10^{8} \ K$). The qualitative 2D evolution up to the runaway is the same as for model A. At time  $\approx9.4$ s, a high pressure region emerges and develop to a detonation within  $\approx0.1$ s (Figs.~\ref{fig:HEL_C1_E_05F_cut_p_all} and .~\ref{fig:HEL_C1_E_05F_cut_p_t9_49}). 
  Since the densities in the lighter envelope of models of type B are lower the burning is milder and the  pressure wave develops into a detonation only in the helium envelope. The ingoing pressure wave burns some of the CO matter, but it is not strong enough to become a detonation. In Fig.~\ref{fig:HEL_C1_E_05F_t9_6_cut_short}, we present pressure, temperature, burning rate and abundance profile along a radial cut in the detonated region at time $t=9.6$ s. It is clearly seen that the pressure wave advancing into the helium (right-hand side) is accompanied by a burning front (red). In the inner side, facing the CO core, we observe only a pressure wave, and the burning rates are negligible. Since our computational region is small, we had to stop the simulation at this stage to avoid artificial boundary effects but it seems that for the less massive envelope of models of type B the detonation develops only in the helium. The burning products here are mainly \iso{40}Ca and \iso{44}Ti. This model run with the \textsc{rich} code gives similar results, a detonation of the Helium shell but no detonation of the CO core.

\
\newpage

\clearpage

\section{Summary and Discussion}

In this study we calculate, for the first time, the crucial stage of the ignition of a detonation in helium envelopes accreted, on top of CO WDs. Helium detonation is an essential first step for any sub-Chandrasekhar SNIa model. In spite of the crucial importance of this stage, all published models up to date ignite the detonation manually and study the subsequent hydrodynamic evolution of the SN. The main difficulties that face the study of the ignition process are as follows:

  \begin{enumerate}[label=(\alph*)]
   \item Due to the extreme temperature sensitivity of the burning rates the spatial scales involved in the ignition process range from a microscopic scale of a few centimetres to the macroscopic scales of the star's scaleheight. With the existing hydrodynamical solvers it is impossible to resolve the entire range of scales.
   
   \item The accreted helium envelope is unstable to convection on time-scales of days, prior to the runaway. Examination of the relevant time-scales shows that there is a stage where the time-scale hierarchy becomes $\tau_{hyd} < \tau_{burn} < \tau_{conv} $. From there on, one can expect that local temperature fluctuations lead to further decrease in the burning time and a detonation ignites locally. In order to study such fluctuations and their ability to ignite a detonation, a multidimensional approach is needed. The complex convective reactive flow is challenging to any hydrodynamical solver. The combination of incomplete resolution and advective terms can lead to numerical heating by hot ashes from adjacent burning regions, heating that can artificially enhance the local burning and develop into a `numeric' detonation.
  \end{enumerate}
   
     Being aware of the obstacles, we adopted a conservative strategy that included two major ingredients. We achieve consistent results using two state of the art, hydrodynamical solvers (\textsc{vulcan} \citep{Livne93}, \textsc{rich} \citep{RICH15}). The numerical schemes of the two hydrodynamic solvers materialize two logically independent algorithms.
     For each solver we adopted the best known limiters to avoid numeric diffusion, specifically limiters that for other related problems proved to give converged results with limited resolution.

     Our main results are as follows:
     
\begin{enumerate}[label=(\alph*)]
      \item  On top of a $ 1.0 \msun$ CO WD for accreted envelopes with mass $\ge0.05 \msun $, temperature fluctuations in the mixed region between the helium envelope and the CO core develop into a helium detonation. 
      
      \item  We predict that for the $0.1 \msun $ envelope a double detonation is ignited (helium detonation in the envelope and CO detonation penetrating the core).

       \item  We predict that for the $0.05 \msun $ envelope a single detonation is ignited (helium) and only a strong pressure wave penetrates into the CO core.
     
      Both results (b and c) are consistent with 1D Lagrangian models of the initial profiles 2D in which we induced a detonation in the helium envelope above the interface between the CO core and got CO detonation for models of type  A and only a pressure wave for models of type B (for more details see  \cite{LivGla90}).     
       
       \item  In the 13 elements alpha reaction net we use, the post detonation abundances are intermediate mass elements with negligible amounts of iron group elements.
\end{enumerate}

The main insights and issues for further discussion are as follows:

\begin{enumerate}[label=(\alph*)]
\item As stated in the previous sections, prior to the helium ignition, the local convective burning heats small regions of matter to high temperatures that float as plumes in the envelope without any significant change in the pressure profile. These local plumes expand the envelope and lower the density. Therefore, the ignited detonation propagates in a diluted envelope. The final abundance of only intermediate mass elements is probably the outcome of this dilution. Therefore, we have indications that the maximal envelope mass that is possible, in order to avoid the production of heavy elements, can be higher than the masses quoted in previous models.
\item Our model adopts cylindrical symmetry and the reaction net we use is a 13 elements alpha net that does not include the reaction sequence \iso{12}C (p, $\gamma$ ) \iso{13}N ( $\alpha$ ,p) \iso{16}O that enhances the burning dramatically above $~1.0 \times 10^{9} \ K$. In 3D we expect higher temperature fluctuations. Therefore, more realistic  models in 3D with higher burning rates will not alter our major result concerning the ignition of a detonation.
\item This work is limited to the ignition stage of the runaway. It shows for the first time that indeed there are conditions in the helium envelope that lead to the ignition of a detonation wave. The structure of the envelope at ignition differs significantly from all of our 1D models. We therefore cannot at this stage, give any predictions as to the total outcome of the runaway at this stage.
       Full star 2D and 3D hydro models with accurate reaction nets are needed in order to check the consistency with 2D simulations.
       We currently do not have the resources to perform the 3D survey.    
\end{enumerate}

\section*{Acknowledgements}

AY would like to thank Phillip Hopkins for the useful discussions.
Part of the simulation were run on the ICORE computer cluster at the Hebrew University. We thank the referee for a detailed reading of our paper and for the constructive commentary.



\bibliographystyle{mnras}
\bibliography{ms_publish} 

\begin{thebibliography}{}
\makeatletter
\relax
\def\mn@urlcharsother{\let\do\@makeother \do\$\do\&\do\#\do\^\do\_\do\%\do\~}
\def\mn@doi{\begingroup\mn@urlcharsother \@ifnextchar [ {\mn@doi@}
  {\mn@doi@[]}}
\def\mn@doi@[#1]#2{\def\@tempa{#1}\ifx\@tempa\@empty \href
  {http://dx.doi.org/#2} {doi:#2}\else \href {http://dx.doi.org/#2} {#1}\fi
  \endgroup}
\def\mn@eprint#1#2{\mn@eprint@#1:#2::\@nil}
\def\mn@eprint@arXiv#1{\href {http://arxiv.org/abs/#1} {{\tt arXiv:#1}}}
\def\mn@eprint@dblp#1{\href {http://dblp.uni-trier.de/rec/bibtex/#1.xml}
  {dblp:#1}}
\def\mn@eprint@#1:#2:#3:#4\@nil{\def\@tempa {#1}\def\@tempb {#2}\def\@tempc
  {#3}\ifx \@tempc \@empty \let \@tempc \@tempb \let \@tempb \@tempa \fi \ifx
  \@tempb \@empty \def\@tempb {arXiv}\fi \@ifundefined
  {mn@eprint@\@tempb}{\@tempb:\@tempc}{\expandafter \expandafter \csname
  mn@eprint@\@tempb\endcsname \expandafter{\@tempc}}}

\bibitem[\protect\citeauthoryear{{Alexander} \& {Ferguson}}{{Alexander} \&
  {Ferguson}}{1994}]{Alexander94}
{Alexander} D.~R.,  {Ferguson} J.~W.,  1994, \mn@doi [\apj] {10.1086/175039},
  \href {http://adsabs.harvard.edu/abs/1994ApJ...437..879A} {437, 879}

\bibitem[\protect\citeauthoryear{{Bildsten}, {Shen}, {Weinberg}  \&
  {Nelemans}}{{Bildsten} et~al.}{2007}]{Bildsten07}
{Bildsten} L.,  {Shen} K.~J.,  {Weinberg} N.~N.,   {Nelemans} G.,  2007,
  \mn@doi [\apjl] {10.1086/519489}, \href
  {http://adsabs.harvard.edu/abs/2007ApJ...662L..95B} {662, L95}

\bibitem[\protect\citeauthoryear{{Blinnikov} \& {Khokhlov}}{{Blinnikov} \&
  {Khokhlov}}{1987}]{BlinnikovKhokhlov87}
{Blinnikov} S.~I.,  {Khokhlov} A.~M.,  1987, Soviet Astronomy Letters, \href
  {http://adsabs.harvard.edu/abs/1987SvAL...13..364B} {13, 364}

\bibitem[\protect\citeauthoryear{{Drout} et~al.,}{{Drout}
  et~al.}{2013}]{Drout13}
{Drout} M.~R.,  et~al., 2013, \mn@doi [\apj] {10.1088/0004-637X/774/1/58},
  \href {http://adsabs.harvard.edu/abs/2013ApJ...774...58D} {774, 58}

\bibitem[\protect\citeauthoryear{{Fink}, {R{\"o}pke}, {Hillebrandt},
  {Seitenzahl}, {Sim}  \& {Kromer}}{{Fink} et~al.}{2010}]{Fink10}
{Fink} M.,  {R{\"o}pke} F.~K.,  {Hillebrandt} W.,  {Seitenzahl} I.~R.,  {Sim}
  S.~A.,   {Kromer} M.,  2010, \mn@doi [\aap] {10.1051/0004-6361/200913892},
  \href {http://adsabs.harvard.edu/abs/2010A%26A...514A..53F} {514, A53}

\bibitem[\protect\citeauthoryear{{Garc{\'{\i}}a-Senz}, {Bravo}  \&
  {Woosley}}{{Garc{\'{\i}}a-Senz} et~al.}{1999}]{Garc99}
{Garc{\'{\i}}a-Senz} D.,  {Bravo} E.,   {Woosley} S.~E.,  1999, \aap, \href
  {http://adsabs.harvard.edu/abs/1999A%26A...349..177G} {349, 177}

\bibitem[\protect\citeauthoryear{Gopala \& van Wachem}{Gopala \& van
  Wachem}{2008}]{Gopala08}
Gopala V.~R.,  van Wachem B.~G.,  2008, \mn@doi [Chemical Engineering Journal]
  {https://doi.org/10.1016/j.cej.2007.12.035}, 141, 204

\bibitem[\protect\citeauthoryear{{Hoeflich} \& {Khokhlov}}{{Hoeflich} \&
  {Khokhlov}}{1996a}]{HoeflichKhokhlov96}
{Hoeflich} P.,  {Khokhlov} A.,  1996a, \mn@doi [\apj] {10.1086/176748}, \href
  {http://adsabs.harvard.edu/abs/1996ApJ...457..500H} {457, 500}

\bibitem[\protect\citeauthoryear{{Hoeflich} \& {Khokhlov}}{{Hoeflich} \&
  {Khokhlov}}{1996b}]{Hirt81}
{Hoeflich} P.,  {Khokhlov} A.,  1996b, \mn@doi [\apj] {10.1086/176748}, \href
  {http://adsabs.harvard.edu/abs/1996ApJ...457..500H} {457, 500}

\bibitem[\protect\citeauthoryear{{Holcomb}, {Guillochon}, {De Colle}  \&
  {Ramirez-Ruiz}}{{Holcomb} et~al.}{2013}]{Holcomb13}
{Holcomb} C.,  {Guillochon} J.,  {De Colle} F.,   {Ramirez-Ruiz} E.,  2013,
  \mn@doi [\apj] {10.1088/0004-637X/771/1/14}, \href
  {http://adsabs.harvard.edu/abs/2013ApJ...771...14H} {771, 14}

\bibitem[\protect\citeauthoryear{Hopkins}{Hopkins}{2015}]{gizmo}
Hopkins P.~F.,  2015, \mn@doi [Mon. Not. Roy. Astron. Soc.]
  {10.1093/mnras/stv195}, 450, 53

\bibitem[\protect\citeauthoryear{{Iglesias} \& {Rogers}}{{Iglesias} \&
  {Rogers}}{1996}]{IGR96}
{Iglesias} C.~A.,  {Rogers} F.~J.,  1996, \mn@doi [\apj] {10.1086/177381},
  \href {http://adsabs.harvard.edu/abs/1996ApJ...464..943I} {464, 943}

\bibitem[\protect\citeauthoryear{{Inserra} et~al.,}{{Inserra}
  et~al.}{2015}]{Insera14}
{Inserra} C.,  et~al., 2015, \mn@doi [\apjl] {10.1088/2041-8205/799/1/L2},
  \href {http://adsabs.harvard.edu/abs/2015ApJ...799L...2I} {799, L2}

\bibitem[\protect\citeauthoryear{{Itoh}, {Mitake}, {Iyetomi}  \&
  {Ichimaru}}{{Itoh} et~al.}{1983}]{Itoh83}
{Itoh} N.,  {Mitake} S.,  {Iyetomi} H.,   {Ichimaru} S.,  1983, \mn@doi [\apj]
  {10.1086/161412}, \href {http://adsabs.harvard.edu/abs/1983ApJ...273..774I}
  {273, 774}

\bibitem[\protect\citeauthoryear{{Jacobs}, {Zingale}, {Nonaka}, {Almgren}  \&
  {Bell}}{{Jacobs} et~al.}{2016}]{Jacobs16}
{Jacobs} A.~M.,  {Zingale} M.,  {Nonaka} A.,  {Almgren} A.~S.,   {Bell} J.~B.,
  2016, \mn@doi [\apj] {10.3847/0004-637X/827/1/84}, \href
  {http://adsabs.harvard.edu/abs/2016ApJ...827...84J} {827, 84}

\bibitem[\protect\citeauthoryear{{Kasliwal} et~al.,}{{Kasliwal}
  et~al.}{2010}]{Kasliwal10}
{Kasliwal} M.~M.,  et~al., 2010, \mn@doi [\apjl] {10.1088/2041-8205/723/1/L98},
  \href {http://adsabs.harvard.edu/abs/2010ApJ...723L..98K} {723, L98}

\bibitem[\protect\citeauthoryear{{Kromer}, {Sim}, {Fink}, {R{\"o}pke},
  {Seitenzahl}  \& {Hillebrandt}}{{Kromer} et~al.}{2010}]{Kromer10}
{Kromer} M.,  {Sim} S.~A.,  {Fink} M.,  {R{\"o}pke} F.~K.,  {Seitenzahl} I.~R.,
    {Hillebrandt} W.,  2010, \mn@doi [\apj] {10.1088/0004-637X/719/2/1067},
  \href {http://adsabs.harvard.edu/abs/2010ApJ...719.1067K} {719, 1067}

\bibitem[\protect\citeauthoryear{{Kushnir}, {Katz}, {Dong}, {Livne}  \&
  {Fern{\'a}ndez}}{{Kushnir} et~al.}{2013}]{Kush13}
{Kushnir} D.,  {Katz} B.,  {Dong} S.,  {Livne} E.,   {Fern{\'a}ndez} R.,  2013,
  \mn@doi [\apjl] {10.1088/2041-8205/778/2/L37}, \href
  {http://adsabs.harvard.edu/abs/2013ApJ...778L..37K} {778, L37}

\bibitem[\protect\citeauthoryear{{Livne}}{{Livne}}{1990}]{Livne90}
{Livne} E.,  1990, \mn@doi [\apjl] {10.1086/185721}, \href
  {http://adsabs.harvard.edu/abs/1990ApJ...354L..53L} {354, L53}

\bibitem[\protect\citeauthoryear{{Livne}}{{Livne}}{1993}]{Livne93}
{Livne} E.,  1993, \mn@doi [\apj] {10.1086/172950}, \href
  {http://adsabs.harvard.edu/abs/1993ApJ...412..634L} {412, 634}

\bibitem[\protect\citeauthoryear{{Livne} \& {Glasner}}{{Livne} \&
  {Glasner}}{1990}]{LivGla90}
{Livne} E.,  {Glasner} A.~S.,  1990, \mn@doi [\apj] {10.1086/169189}, \href
  {http://adsabs.harvard.edu/abs/1990ApJ...361..244L} {361, 244}

\bibitem[\protect\citeauthoryear{{Livne} \& {Glasner}}{{Livne} \&
  {Glasner}}{1991}]{LivGla91}
{Livne} E.,  {Glasner} A.~S.,  1991, \mn@doi [\apj] {10.1086/169813}, \href
  {http://adsabs.harvard.edu/abs/1991ApJ...370..272L} {370, 272}

\bibitem[\protect\citeauthoryear{{Moll} \& {Woosley}}{{Moll} \&
  {Woosley}}{2013}]{MollWoosley13}
{Moll} R.,  {Woosley} S.~E.,  2013, \mn@doi [\apj]
  {10.1088/0004-637X/774/2/137}, \href
  {http://adsabs.harvard.edu/abs/2013ApJ...774..137M} {774, 137}

\bibitem[\protect\citeauthoryear{{Nomoto}}{{Nomoto}}{1980}]{Nomoto80}
{Nomoto} K.,  1980, \mn@doi [\ssr] {10.1007/BF00168350}, \href
  {http://adsabs.harvard.edu/abs/1980SSRv...27..563N} {27, 563}

\bibitem[\protect\citeauthoryear{{Nomoto}}{{Nomoto}}{1982}]{Nomoto82}
{Nomoto} K.,  1982, \mn@doi [\apj] {10.1086/160031}, \href
  {http://adsabs.harvard.edu/abs/1982ApJ...257..780N} {257, 780}

\bibitem[\protect\citeauthoryear{{Nugent}, {Baron}, {Branch}, {Fisher}  \&
  {Hauschildt}}{{Nugent} et~al.}{1997}]{Nugent97}
{Nugent} P.,  {Baron} E.,  {Branch} D.,  {Fisher} A.,   {Hauschildt} P.~H.,
  1997, \mn@doi [\apj] {10.1086/304459}, \href
  {http://adsabs.harvard.edu/abs/1997ApJ...485..812N} {485, 812}

\bibitem[\protect\citeauthoryear{{Perets}, {Badenes}, {Arcavi}, {Simon}  \&
  {Gal-yam}}{{Perets} et~al.}{2011}]{Perets11}
{Perets} H.~B.,  {Badenes} C.,  {Arcavi} I.,  {Simon} J.~D.,   {Gal-yam} A.,
  2011, \mn@doi [\apj] {10.1088/0004-637X/730/2/89}, \href
  {http://adsabs.harvard.edu/abs/2011ApJ...730...89P} {730, 89}

\bibitem[\protect\citeauthoryear{{Poznanski} et~al.,}{{Poznanski}
  et~al.}{2010}]{Poznanski10}
{Poznanski} D.,  et~al., 2010, \mn@doi [Science] {10.1126/science.1181709},
  \href {http://adsabs.harvard.edu/abs/2010Sci...327...58P} {327, 58}

\bibitem[\protect\citeauthoryear{{Shen} \& {Bildsten}}{{Shen} \&
  {Bildsten}}{2009}]{ShenBild09}
{Shen} K.~J.,  {Bildsten} L.,  2009, \mn@doi [\apj]
  {10.1088/0004-637X/699/2/1365}, \href
  {http://adsabs.harvard.edu/abs/2009ApJ...699.1365S} {699, 1365}

\bibitem[\protect\citeauthoryear{{Shen} \& {Moore}}{{Shen} \&
  {Moore}}{2014}]{ShenMoore14}
{Shen} K.~J.,  {Moore} K.,  2014, \mn@doi [\apj] {10.1088/0004-637X/797/1/46},
  \href {http://adsabs.harvard.edu/abs/2014ApJ...797...46S} {797, 46}

\bibitem[\protect\citeauthoryear{{Shen}, {Kasen}, {Weinberg}, {Bildsten}  \&
  {Scannapieco}}{{Shen} et~al.}{2010}]{Shen10}
{Shen} K.~J.,  {Kasen} D.,  {Weinberg} N.~N.,  {Bildsten} L.,   {Scannapieco}
  E.,  2010, \mn@doi [\apj] {10.1088/0004-637X/715/2/767}, \href
  {http://adsabs.harvard.edu/abs/2010ApJ...715..767S} {715, 767}

\bibitem[\protect\citeauthoryear{{Sim}, {Fink}, {Kromer}, {R{\"o}pke}, {Ruiter}
   \& {Hillebrandt}}{{Sim} et~al.}{2012}]{Sim12}
{Sim} S.~A.,  {Fink} M.,  {Kromer} M.,  {R{\"o}pke} F.~K.,  {Ruiter} A.~J.,
  {Hillebrandt} W.,  2012, \mn@doi [\mnras] {10.1111/j.1365-2966.2011.20162.x},
  \href {http://adsabs.harvard.edu/abs/2012MNRAS.420.3003S} {420, 3003}

\bibitem[\protect\citeauthoryear{{Spiegel}}{{Spiegel}}{1963}]{spi63}
{Spiegel} E.~A.,  1963, \mn@doi [\apj] {10.1086/147628}, \href
  {http://adsabs.harvard.edu/abs/1963ApJ...138..216S} {138, 216}

\bibitem[\protect\citeauthoryear{{Springel}}{{Springel}}{2010}]{arepo}
{Springel} V.,  2010, \mn@doi [\mnras] {10.1111/j.1365-2966.2009.15715.x},
  \href {http://adsabs.harvard.edu/abs/2010MNRAS.401..791S} {401, 791}

\bibitem[\protect\citeauthoryear{{Taam}}{{Taam}}{1980}]{Taam80}
{Taam} R.~E.,  1980, \mn@doi [\apj] {10.1086/157852}, \href
  {http://adsabs.harvard.edu/abs/1980ApJ...237..142T} {237, 142}

\bibitem[\protect\citeauthoryear{{Townsley}, {Moore}  \& {Bildsten}}{{Townsley}
  et~al.}{2012}]{Townsley12}
{Townsley} D.~M.,  {Moore} K.,   {Bildsten} L.,  2012, \mn@doi [\apj]
  {10.1088/0004-637X/755/1/4}, \href
  {http://adsabs.harvard.edu/abs/2012ApJ...755....4T} {755, 4}

\bibitem[\protect\citeauthoryear{{Ubbink} \& {Issa}}{{Ubbink} \&
  {Issa}}{1999}]{Ubbnik99}
{Ubbink} O.,  {Issa} R.~I.,  1999, \mn@doi [Journal of Computational Physics]
  {10.1006/jcph.1999.6276}, \href
  {http://adsabs.harvard.edu/abs/1999JCoPh.153...26U} {153, 26}

\bibitem[\protect\citeauthoryear{{Waldman}, {Sauer}, {Livne}, {Perets},
  {Glasner}, {Mazzali}, {Truran}  \& {Gal-Yam}}{{Waldman}
  et~al.}{2011}]{Waldman11}
{Waldman} R.,  {Sauer} D.,  {Livne} E.,  {Perets} H.,  {Glasner} A.,  {Mazzali}
  P.,  {Truran} J.~W.,   {Gal-Yam} A.,  2011, \mn@doi [\apj]
  {10.1088/0004-637X/738/1/21}, \href
  {http://adsabs.harvard.edu/abs/2011ApJ...738...21W} {738, 21}

\bibitem[\protect\citeauthoryear{{Woosley} \& {Kasen}}{{Woosley} \&
  {Kasen}}{2011}]{WoosleyKasen11}
{Woosley} S.~E.,  {Kasen} D.,  2011, \mn@doi [\apj]
  {10.1088/0004-637X/734/1/38}, \href
  {http://adsabs.harvard.edu/abs/2011ApJ...734...38W} {734, 38}

\bibitem[\protect\citeauthoryear{{Woosley} \& {Weaver}}{{Woosley} \&
  {Weaver}}{1994}]{WoosleyWeaver94}
{Woosley} S.~E.,  {Weaver} T.~A.,  1994, \mn@doi [\apj] {10.1086/173813}, \href
  {http://adsabs.harvard.edu/abs/1994ApJ...423..371W} {423, 371}

\bibitem[\protect\citeauthoryear{{Woosley}, {Taam}  \& {Weaver}}{{Woosley}
  et~al.}{1986}]{Woosley86}
{Woosley} S.~E.,  {Taam} R.~E.,   {Weaver} T.~A.,  1986, \mn@doi [\apj]
  {10.1086/163926}, \href {http://adsabs.harvard.edu/abs/1986ApJ...301..601W}
  {301, 601}

\bibitem[\protect\citeauthoryear{{Yalinewich}, {Steinberg}  \&
  {Sari}}{{Yalinewich} et~al.}{2015}]{RICH15}
{Yalinewich} A.,  {Steinberg} E.,   {Sari} R.,  2015, \mn@doi [\apjs]
  {10.1088/0067-0049/216/2/35}, \href
  {http://adsabs.harvard.edu/abs/2015ApJS..216...35Y} {216, 35}

\bibitem[\protect\citeauthoryear{{Zel'Dovich}, {Librovich}, {Makhviladze}  \&
  {Sivashinskil}}{{Zel'Dovich} et~al.}{1970}]{Zel70}
{Zel'Dovich} Y.~B.,  {Librovich} V.~B.,  {Makhviladze} G.~M.,   {Sivashinskil}
  G.~I.,  1970, \mn@doi [Journal of Applied Mechanics and Technical Physics]
  {10.1007/BF00908106}, \href
  {http://adsabs.harvard.edu/abs/1970JAMTP..11..264Z} {11, 264}

\bibitem[\protect\citeauthoryear{{Zingale}, {Nonaka}, {Almgren}, {Bell},
  {Malone}  \& {Orvedahl}}{{Zingale} et~al.}{2013}]{Zingale13}
{Zingale} M.,  {Nonaka} A.,  {Almgren} A.~S.,  {Bell} J.~B.,  {Malone} C.~M.,
  {Orvedahl} R.~J.,  2013, \mn@doi [\apj] {10.1088/0004-637X/764/1/97}, \href
  {http://adsabs.harvard.edu/abs/2013ApJ...764...97Z} {764, 97}

\makeatother
\end{thebibliography}

\bsp	
\label{lastpage}
\end{document}